\documentclass[letterpaper,english,aps,prb,reprint,amsmath,amssymb,showpacs]{revtex4-1}
\usepackage[T1]{fontenc}
\usepackage[latin9]{inputenc}
\usepackage{color}
\usepackage{babel}
\usepackage{bm}
\usepackage{amsmath}
\usepackage{amssymb}
\usepackage{graphicx}
\usepackage{esint}
\usepackage[unicode=true,pdfusetitle,
 bookmarks=true,bookmarksnumbered=true,bookmarksopen=true,bookmarksopenlevel=1,
 breaklinks=false,pdfborder={0 0 0},pdfborderstyle={},backref=false,colorlinks=true]
 {hyperref}

\makeatletter

\pdfpageheight\paperheight
\pdfpagewidth\paperwidth

\usepackage{braket}
\usepackage{mathtools}
\usepackage{microtype}
\usepackage{graphicx}
\hypersetup{linkcolor = blue,
            urlcolor  = blue,
            citecolor = blue,
            anchorcolor = blue}

\makeatother

\begin{document}

\title{Intermolecular mechanism for multiple maxima in molecular dynamic susceptibility}

\author{Le Tuan Anh Ho}
\email{anh.holetuan@chem.kuleuven.be }

\affiliation{Theory of Nanomaterials Group, Katholieke Universiteit Leuven, Celestijnenlaan 200F, B-3001 Leuven, Belgium}

\affiliation{Institute of Research and Development, Duy Tan University, Da Nang, Viet Nam}

\author{Liviu F. Chibotaru}
\email{liviu.chibotaru@chem.kuleuven.be }

\affiliation{Theory of Nanomaterials Group, Katholieke Universiteit Leuven, Celestijnenlaan 200F, B-3001 Leuven, Belgium}

\date{\today}
\begin{abstract}
A simple two-level system is introduced to demonstrate the existence of the intermolecular mechanism of the appearance/disappearance of multiple maxima in molecular dynamic susceptibility at low temperature. With minimum two relaxation processes, quantum tunneling induced by the nuclear spin bath and the direct process due to the spin-phonon interaction, we prove that the existence of a sufficiently wide dipolar field distribution in the experimental sample is the main cause of this phenomenon in zero or weak applied dc field.  The correlations between the phenomenon and the applied dc field, temperature, or magnetic dilution of the sample are also investigated. Application to several experimental systems has shown a good agreement between the proposed theory and experiments. Cases with more complex multiplet spectrum, more relaxation processes, and possibility of more maxima in the molecular dynamic susceptibility are discussed. 
\end{abstract}

\maketitle

\section{Introduction}

\global\long\def\hmt{\mathcal{H}}
\global\long\def\vt#1{\mathbf{#1}}

\global\long\def\chip{\chi'}
\global\long\def\chipp{\chi''}

Owning to a slow magnetic relaxation manifested, single-molecule magnets (SMMs) have attracted an increasing attention in chemistry, physics, and material science due to prospects for applications in high density magnetic storage, quantum information, and spintronic devices \cite{Leuenberger2001,Hill2003,Vincent2012,Bogani2008,Gatteschi2003,Sessoli1993,Tejada2001,Wernsdorfer2002,Godfrin2017,Moreno-Pineda2017}. In order to characterize a SMM complex, one of the common procedures is measuring its relaxation time via the response to an external ac magnetic field. The obtained ac susceptibility data is then used to extract the relaxation time $\tau$ via the frequency location $\omega_{max}$ of the maximum of out-of-phase susceptibility $\chipp\left(\omega\right)$, $\tau=\omega_{max}^{-1}$. However, this conventional method of the relaxation time extraction recently faces a challenge due to numerous observations of multiple maxima in $\chipp\left(\omega\right)$ in both polynuclear and mononuclear SMMs recently \cite{Boca2014a,Arauzo2014,Rajnak2017,Guo2010,Amjad2016,Li2016,Holmberg2016,Holmberg2015,Boca2017,Peng2017a,Habib2015,Prodius2013,Rinehart2010,Jeletic2011,Lucaccini2016b,Ruiz2012,Li2015b,Rajnak2014a,AlHareri2016,Miklovic2015,Jiang2011,Gregson2015,Cosquer2013,Zadrozny2011a,Huang2013,Meihaus2011,Liu2013a}. For polynuclear complexes, this phenomenon is often reckoned as originating from distinct relaxation pathways of different kinds/structures of magnetic ions in the compounds \cite{Blagg2013b,Zadrozny2011a,Hewitt2010a,Lin2009a,Guo2011d,Amjad2016,Guo2011a,Guo2010,Hewitt2009}. Another intramolecular mechanism due to the existence of several relaxation modes has also been proposed to address this phenomenon in both mononuclear and polynuclear compounds within a three-level model \cite{Ho2016}. However, both mechanisms suffer from some weaknesses. In particular, while the former is unable to explain the phenomenon in mononuclear SMMs, the latter cannot provide a good description at sufficiently low temperature where only two energy levels are effectively populated, i.e., when the temperature is much lower than the blocking barrier of the SMM. 

On the other hand, several recent experiments on mononuclear SMMs have also shown that the dilution of the SMM samples may lead to an appearance/disappearance of two maxima in the out-of-phase susceptibility \cite{Li2015b,Habib2015,Li2016,Huang2013}. This gave rise to a speculation that the intermolecular interaction in the crystal mediates the phenomenon \cite{Habib2015,Li2016,Huang2013}. Additionally, temperature or applied dc field were also shown to have strong effects on the phenomenon. Importantly, it is found in some of these works that the extracted energy barrier is much higher than the temperature at which the phenomenon occurs. These rule out the two earlier proposed mechanisms. Another mechanism is hence required to explain the phenomenon as well as its dependence on the applied dc field, temperature, and dilution which are not fully covered by the previous mechanisms.

In this work, by employing a minimum two-level model with two relaxation processes, namely quantum tunneling induced by nuclear spins and direct process (Raman process can be trivially added into the model), we elucidate the appearance/disappearance of two maxima in the out-of-phase susceptibility of various types of SMM samples. We prove that a wide dipolar field distribution can indeed give rise to the phenomenon in low temperature regime. Effects of applied dc field, temperature, and dilution are also clarified. The mechanism is generic and relevant for both single-crystal and polycrystalline (powder) samples, as well as applicable for both mononuclear and polynuclear SMM compounds. 

The article is organized as follows. In Section II, the microscopic description of dynamic susceptibility in an SMM system at low temperature is introduced. Using this section's results, we study the appearance/disappearance of two maxima in the out-of-phase susceptibility in both non-Kramers and Kramers SMM systems without and with the presence of the intermolecular interaction in Section III and IV, respectively. Section V is then dedicated to an investigation of the effect of the applied dc field, temperature, and dilution to the appearance/disappearance of the multiple maxima in the dynamic susceptibility as well as applications of the proposed theory to several experimental systems. Discussions are given in the last section.

\section{Microscopic description of dynamic susceptibility in single-molecule magnets at low temperature \label{sec:Microscopic-description}}

A single-molecule magnet made of magnetic ion(s) and ligands, characterized by a large spin number $S$ $\left(J\right)$, in weak interaction with a thermal bath at sufficiently low temperature so that it can be considered as a two-level system is studied. Residing in a crystal, this \emph{central} molecular spin is also interacting with nuclear spins and other molecules in the vicinity. Since the energy splitting caused by the local magnetic field at the nuclear sites is often much smaller than the temperature at which experiments are conducted (at the order of Kelvin), it is supposed that every nuclear spin is in a completely disordered equilibrium states. This allows us to treat the surrounding nuclear spins like a spin bath. The dipolar effect from other SMM molecules in the crystal, on the other hand, is modeled in a classical way by a local mean field at the central spin's site for simplicity. Since the ac susceptibility of a magnetic sample is measured in the thermal equilibrium state, in a zero or weak applied field, the local dipolar mean field then can be statistically treated as a Gaussian probability distribution \cite{Berkov1996} along three reference frame axes $\alpha=x,y,z$, with the same standard deviation $H_{dm}$, i.e., 
\begin{equation}
p\left(H_{d,\alpha}\right)=\frac{1}{\sqrt{2\pi}H_{dm}}e^{-H_{d,\alpha}^{2}/2H_{dm}^{2}}.\label{eq:p(H_d,a)}
\end{equation}

Finding the linear response of the central spin system in interaction with numerous surrounding spins at a first glance seems like a daunting task. However, this problem can be phenomenologically treated using a specific rate equation in combination with an interesting relaxation property figured out by A. Vijayaraghavan and A. Garg \cite{Vijayaraghavan2009,Vijayaraghavan2011}. In particular, it was found that a central spin surrounding by a molecular spin environment and in interaction with a nuclear spin bath will experience an \emph{incoherent} relaxation where the nuclear spin bath plays the primary role in decoherence process meanwhile the molecular spin environment creates the energy bias. Making use of these, we can write a simple rate equation for the diagonal density matrix elements in the eigenstates basis of the pseudospin $\tilde{S}_{z}=1/2$ of the central non-Kramers/Kramers doublet\cite{Vijayaraghavan2009,Vijayaraghavan2011,Griffith1963a,Ho2017,Chibotaru2014a}: 
\begin{equation}
\frac{d\rho_{m}}{dt}=\Gamma_{mm'}\rho_{m'}-\Gamma_{m'm}\rho_{m},\label{eq:rate-equation}
\end{equation}
where $\ket{m}$ and $\ket{m'}$ are eigenvectors of the pseudospin doublet $\tilde{S}_{z}$ corresponding to the ground doublet of the central spin, and $\Gamma_{mm'}$ $\left(\Gamma_{m'm}\right)$ consists of contributions from the quantum tunneling process induced by the nuclear bath and the direct process induced by the phonon (thermal) bath: 
\begin{gather}
\Gamma_{mm'}=\Gamma_{tn}+\Gamma_{dr}\approx\Gamma_{m'm}/c,\,c\approx e^{W/kT},\label{eq:Gamma_mm'-1}\\
\Gamma_{tn}=\frac{\sqrt{2\pi}}{4}\frac{\Delta^{2}}{W_{n}}e^{-W^{2}/2W_{n}^{2}},\label{eq:Gamma_mm'-2}\\
\Gamma_{dr}\approx\frac{C_{dr}W^{3}}{c-1},\label{eq:Gamma_mm'-3}
\end{gather}
where $W_{n}$ characterizes the nuclear spin bath effect and is of order of magnitude of the collective magnetic field caused by nuclear spins at the central spin location, $\Delta$ is the tunneling splitting of the central spin ground doublet, and $W$ is the energy bias between $\ket{m}$ and $\ket{m'}$ caused by the external and dipolar field \cite{Vijayaraghavan2009,Vijayaraghavan2011}. In accordance with Ref. [\onlinecite{Vijayaraghavan2011}], we have also assumed that  $\Delta\ll W_{n}\ll kT$. Additionally, the direct process is supposed to be negligible compared to the quantum tunneling process at resonance. It should also be noted that the factor $C_{dr}$ is a constant in non-Kramers systems but field-dependent in Kramers systems \cite{Abragam1970}. Orbach relaxation process, which is insignificant at low temperature and well-separated excited doublets, is omitted. Furthermore, the phonon-induced incoherent quantum tunneling process is supposed to be small comparing to $\Gamma_{tn}$ as the thermal escape rate to excited states is negligible in this case. For simplicity, in the following mechanism demonstration, Raman process is not included. However, this relaxation process can be trivially added in the expression of $\Gamma_{mm'}$ (as seen in Sec. \ref{sec:Theory vs. experiment}), Eq. \eqref{eq:Gamma_mm'-1}, without any effect to the mechanism.

In order to simulate the ac susceptibility measurement, an external magnetic field consisting of a static and a small oscillating component $\vt H_{ext}=\vt H_{dc}+\vt h\cos\omega t$ is applied. Denoting $\left(\theta,\varphi\right)$ respectively as the azimuthal and polar angle between $\vt H_{ext}$ and the main magnetic axes of the molecular central spin, the Hamiltonian for the ground doublet of the central spin system can be written as follows : 
\begin{multline}
\hmt=\frac{\Delta}{2}\left(\ket{m}\bra{m'}+\ket{m'}\bra{m}\right)\\
+\frac{W+V\cos\omega t}{2}\left(\ket{m}\bra{m}-\ket{m'}\bra{m'}\right),\label{eq:Hamiltonian}
\end{multline}
where 
\begin{gather}
W=-2\mu_{z,m}\left(H_{dc}\cos\theta+H_{d,z}\right),\\
V=-2\mu_{z,m}h\cos\theta.
\end{gather}
Here $\mu_{z,m}$ is the expectation value of the central spin magnetic moment component $\mu_{z}$ in the state $\ket{m}$, $\Delta=\sqrt{g_{x}^{2}\left(H_{dc,x}+H_{d,x}\right)^{2}+g_{y}^{2}\left(H_{dc,y}+H_{d,y}\right)^{2}}$ for Kramers systems or intrinsic for non-Kramers systems. Here we have used the property of the time-odd operator $\bm{\mu}_{m'}=-\bm{\mu}_{m}$ and the fact that the transverse components of $\bm{\mu}$ are negligible as the spin number $S$ of the central spin is large.

From the rate equation \eqref{eq:rate-equation} and the Hamiltonian \eqref{eq:Hamiltonian}, linear response of the diagonal density matrix elements $\delta\bm{\rho}=\left(\delta\rho_{m},\delta\rho_{m'}\right)$ to the small harmonic perturbation $V\cos\omega t$ can be easily found \cite{Garanin2011,Ho2016}: 
\begin{equation}
\delta\bm{\rho}=2\left(\frac{\lambda}{\lambda^{2}+\omega^{2}}\cos\omega t+\frac{\omega}{\lambda^{2}+\omega^{2}}\sin\omega t\right)\frac{\left(\mathbf{L}\cdot\mathbf{f}\right)}{\left(\mathbf{L}\cdot\mathbf{R}\right)}\mathbf{R},
\end{equation}
where $\mathbf{L}=\left(-c,1\right)$ and $\mathbf{R}=\left(-1,1\right)$ are respectively left- and right-eigenvector corresponding to the non-zero eigenvalue $\lambda=\left(1+c\right)\Gamma_{mm'}$ of the relaxation matrix $\Phi=\Gamma_{mm'}\begin{pmatrix}-c & 1\\
c & -1
\end{pmatrix}$, and $\mathbf{f}=\frac{c}{c+1}\frac{\mu_{z,m}h\cos\theta}{kT}\Gamma_{mm'}\left(1,-1\right)$, where $c$ is defined in Eq. \eqref{eq:Gamma_mm'-1}. Accordingly, linear response of the magnetization of a crystal along the direction of the applied field as measured in ac susceptibility $\delta M=h\left(\chip\cos\omega t+\chipp\sin\omega t\right)\approx n\left(\bm{\mu}_{z}\cdot\delta\bm{\rho}\right)\cos\theta$ can be easily calculated: 
\begin{gather}
\chi'=4\chi_{0}\frac{c\Gamma_{mm'}^{2}\cos^{2}\theta}{\left(1+c\right)^{2}\Gamma_{mm'}^{2}+\omega^{2}},\label{eq:crystal in-phase susceptibility - no internal field}\\
\chi''=4\chi_{0}\frac{c}{c+1}\frac{\omega\Gamma_{mm'}\cos^{2}\theta}{\left(1+c\right)^{2}\Gamma_{mm'}^{2}+\omega^{2}},\label{eq:crystal out-of-phase susceptibility - no internal field}
\end{gather}
where $\chi_{0}\equiv n\mu_{z,m}^{2}/kT$ and $n$ is the volume/molar molecular spin density of the crystal. 

Averaging over directions of applied field of the expressions \eqref{eq:crystal in-phase susceptibility - no internal field} and \eqref{eq:crystal out-of-phase susceptibility - no internal field} gives the in-phase $\left\langle \chi'\right\rangle $ and out-of-phase susceptibility $\left\langle \chi''\right\rangle $ of a polycrystalline (powder) sample, 
\begin{gather}
\left\langle \chi'\right\rangle =\frac{1}{4\pi}\int d\varphi\,d\theta\,\sin\theta\chip,\label{eq:powder in-phase susceptibility - general case}\\
\left\langle \chi''\right\rangle =\frac{1}{4\pi}\int d\varphi\,d\theta\,\sin\theta\chipp.\label{eq:powder out-of-phase susceptibility - general case}
\end{gather}

Below we will use these expressions to investigate the possibility of multiple maxima in $\chipp\left(w\right)$ under different conditions in single-crystal and polycrystalline samples. For simplicity in notation, hereinafter, $\chi_{0}$ will be used as unit of in-phase and out-of-phase susceptibility.

\section{Low-temperature dynamic susceptibility in zero internal dipolar field\label{sec:No dipolar field}}

From Eq. \eqref{eq:crystal out-of-phase susceptibility - no internal field}, it is easy to see that in a constant external field, $\chipp$ of a single-crystal sample can only have one maximum at the frequency $\omega_{\mathrm{max}}=\Gamma_{\mathrm{relax}}\equiv\left(1+c\right)\Gamma_{mm'}$. However, this may be invalid for polycrystalline samples where the variation of the orientation of the applied dc field relative to the main anisotropy axis of a molecule induces a change of relaxation time in the latter, which gives rise to a distribution in the relaxation times. This situation looks similar to the appearance of distinct EPR signals corresponding to different $g$-factors (e.g., $g_{\parallel}$ and $g_{\perp}$) in powder samples. This thus inspire us to investigate the relevance of this ``powder EPR'' mechanism for the arising of two maxima in the out-of-phase susceptibility $\chipp$.

\subsection{Non-Kramers systems \label{subsec:No dipolar field - non-Kramers}}

In non-Kramers system, transverse components of the magnetic field hardly affect the tunneling splitting $\Delta$. Consequently, we can ignore them in the expressions of the dynamic susceptibility. Substituting Eqs. (\ref{eq:Gamma_mm'-1}-\ref{eq:Gamma_mm'-3}) for $\Gamma_{mm'}$ into Eqs. \eqref{eq:powder in-phase susceptibility - general case} and \eqref{eq:powder out-of-phase susceptibility - general case}, and defining the following quantities: 
\begin{gather}
H_{n}\equiv-W_{n}/2\mu_{z,m},\,h_{dc}\equiv H_{dc}/H_{n},\\
\Gamma_{tn,0}\equiv\frac{\sqrt{2\pi}}{4}\frac{\Delta^{2}}{W_{n}},\,\Gamma_{dr,0}\equiv C_{dr}W_{n}^{2}kT,\label{eq:Gamma_tn0-Gamma_dr0}\\
\alpha\equiv\sqrt{\Gamma_{dr,0}/\Gamma_{tn,0}},\,w\equiv\omega/2\Gamma_{tn,0},\,x=\cos\theta,
\end{gather}
we obtain the expressions for $\left\langle \chip\right\rangle $ and $\left\langle \chipp\right\rangle $ of a polycrystalline sample in zero internal dipolar field: 
\begin{gather}
\left\langle \chi'\right\rangle =\intop_{0}^{1}dx\frac{\left(e^{-h_{dc}^{2}x^{2}/2}+\alpha^{2}h_{dc}^{2}x^{2}\right)^{2}x^{2}}{\left(e^{-h_{dc}^{2}x^{2}/2}+\alpha^{2}h_{dc}^{2}x^{2}\right)^{2}+w^{2}},\label{eq:powder in-phase susceptibility - no internal field 2-1}\\
\left\langle \chi''\right\rangle =\intop_{0}^{1}dx\frac{w\left(e^{-h_{dc}^{2}x^{2}/2}+\alpha^{2}h_{dc}^{2}x^{2}\right)x^{2}}{\left(e^{-h_{dc}^{2}x^{2}/2}+\alpha^{2}h_{dc}^{2}x^{2}\right)^{2}+w^{2}}.\label{eq:powder out-of-phase susceptibility - no internal field 2-1}
\end{gather}
For simplicity, here we have assumed that $W\ll kT$. Since the above expressions cannot be integrated analytically, we resort to a numerical investigation of the appearance/disappearance of two maxima in the out-of-phase susceptibility. Further, some typical examples with reasonable values of the parameters will be given. 

From Eqs. \eqref{eq:powder in-phase susceptibility - no internal field 2-1} and \eqref{eq:powder out-of-phase susceptibility - no internal field 2-1}, one can see that $\left\langle \chi'\right\rangle $ and $\left\langle \chi''\right\rangle $ are functions of three parameters: $h_{dc}$, $\alpha$, and $w$. Considering that $H_{n}$, the characteristic hyperfine field, is of the order of several tens to hundreds Oersted \cite{Gatteschi2006}, $h_{dc}$ typically runs from 1 to 10. Meanwhile, being strongly dependent on $W_{n}$ and the tunneling splitting $\Delta$, the value range of $\alpha$, Eq. \eqref{eq:Gamma_tn0-Gamma_dr0}, is much broader. Since the ac frequency $f$ in dynamic magnetic measurements is often between $1$ Hz to 1000 Hz \cite{Gatteschi2006,Bartolome2014a}, the dimensionless quantity $w=\pi f/\Gamma_{tn,0}$ also varies over several orders of magnitude.

\begin{figure}
\includegraphics{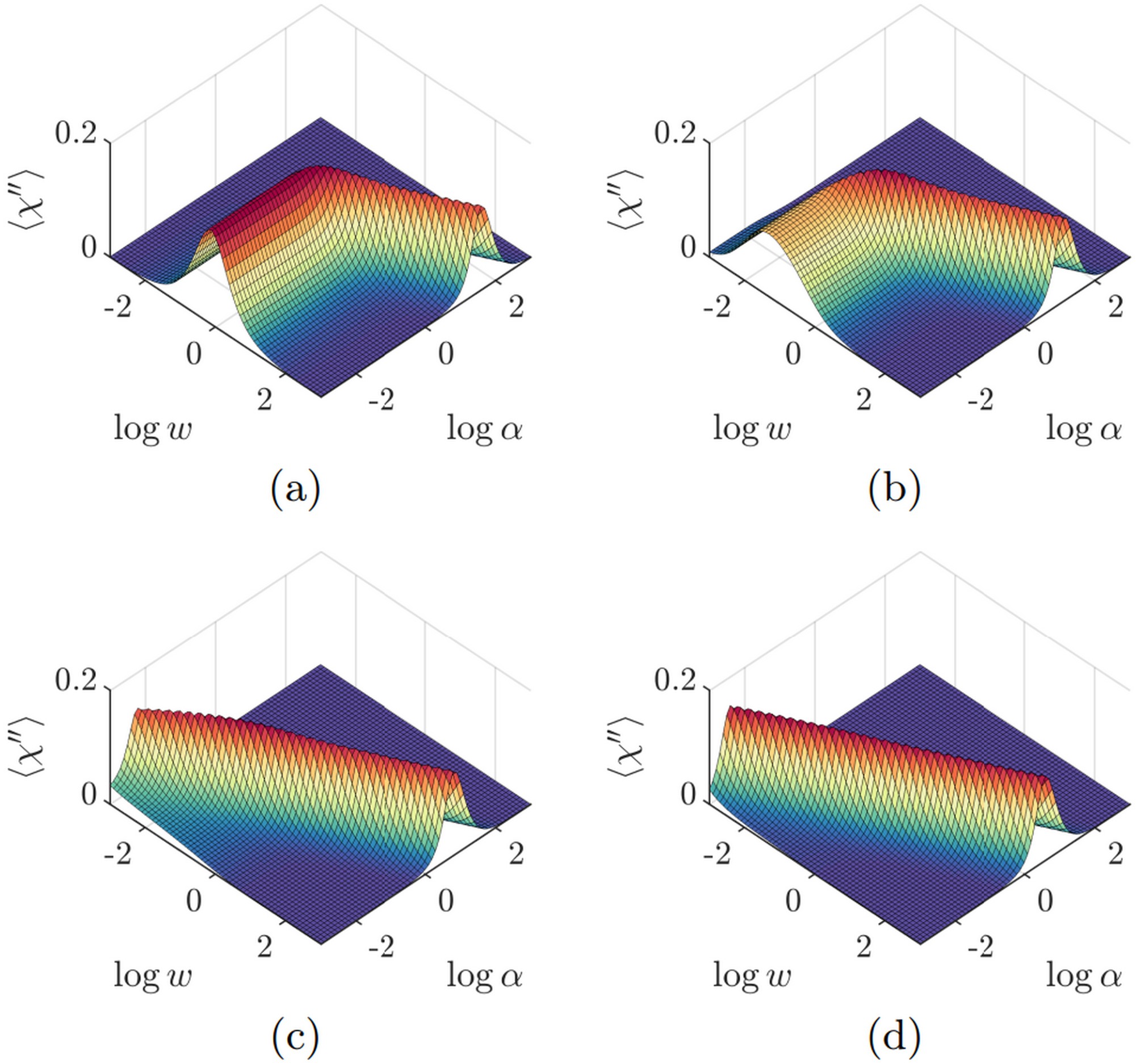}

\caption{Out-of-phase susceptibility of a non-Kramers polycrystalline sample ($\chi_{0}$ units) without an internal dipolar field but with an applied dc field $h_{dc}=1$ (a), 3 (b), 6 (c), and 10 (d).\label{fig:chipp - non-Kramer - no dipolar field}}
\end{figure}

Fig. \ref{fig:chipp - non-Kramer - no dipolar field} shows the variation of the out-of-phase susceptibility versus $\left(\alpha,w\right)$ in the logarithmic scale for several values of $h_{dc}$ in the non-Kramers polycrystalline samples. The corresponding Cole-Cole plot are shown in the Supplemental Material\footnote{See Supplemental Material at {[}URL inserted by the publisher{]} for a general formula of the direct transition rate between two ground doublet's states and more plots of out-of-phase susceptibility w.r.t. frequency and Cole-Cole plots for both single-crystal and polycrystalline non-Kramers and Kramers samples.}. From this figure, it is clear that a spherical uniform distribution of the external field w.r.t. molecular frame cannot give rise to two maxima in $\chipp\left(\omega\right)$. The only result close to the two-maxima situation is a small shoulder shown in Fig. \ref{fig:chipp - non-Kramer - no dipolar field}c for $h_{dc}=6$. 

\subsection{Kramers systems \label{subsec:no dipolar field - Kramers system}}

Contrary to non-Kramers systems, in Kramers molecules the transverse components of the magnetic field play an important role in relaxation. In particular, they open a tunneling splitting gap through which the quantum tunneling of the magnetization takes place. Moreover, they also define the direct relaxation process rate by coupling the ground doublet's states with other excited states. This raises a difficulty in demonstrating the appearance/disappearance of two maxima in Kramers system since the general expression of the direct transition rate between two ground doublet's states contains too many unknown parameters (see Supplemental Material \cite{Note1} for the general formula). For simplicity, we thus resort to a simple approximation which takes into account only the rotational contribution of the crystal lattice deformations\cite{Chudnovsky2005,Calero2006a,Garanin2011,Ho2017}. In this approximation, the direct transition rate from $\ket{m}$ to $\ket{m'}$ in the limit where the system is out of resonance is given by \cite{Ho2017}: 
\begin{equation}
\Gamma_{mm'}\approx CkTW^{4},
\end{equation}
where $C$ is a constant. Here the assumptions $W\ll kT$ have also been used.

In the presence of an applied magnetic field, the tunneling splitting and the bias in the ground doublet are given by: 
\begin{align}
\Delta & =\sqrt{g_{x}^{2}H_{x}^{2}+g_{y}^{2}H_{y}^{2}},\\
W & =g_{z}H_{z},
\end{align}
where $g_{i},i=x,y,z$, are the principal components of the $g$-tensor of the ground doublet described by $\tilde{S}=1/2$ pseudospin \cite{Chibotaru2014a,Ho2017}. Since it is often the case that $g_{x}$ and $g_{y}$ are of the same order of magnitude in SMMs, without loss of generality, we further assume that $g_{x}=g_{y}=g_{\perp}$. This yields 
\begin{equation}
\Gamma_{mm'}=\frac{\sqrt{2\pi}}{4}\frac{g_{\perp}^{2}\left(H_{x}^{2}+H_{y}^{2}\right)}{W_{n}}e^{-g_{z}^{2}H_{z}^{2}/2W_{n}^{2}}+CkTg_{z}^{4}H_{z}^{4}.\label{eq:total relaxation rate - Kramers system}
\end{equation}

Substituting the above expression into Eqs. (\ref{eq:powder in-phase susceptibility - general case}-\ref{eq:powder out-of-phase susceptibility - general case}) and redefining: 
\begin{gather}
H_{n}\equiv W_{n}/g_{z},\\
\Gamma_{tn,0}\equiv\frac{\sqrt{2\pi}}{4}\frac{g_{\perp}^{2}H_{n}}{g_{z}},\Gamma_{dr,0}\equiv CkTg_{z}^{4}H_{n}^{4},\\
\alpha\equiv\sqrt{\Gamma_{dr,0}/\Gamma_{tn,0}},
\end{gather}
we obtain the expressions for $\left\langle \chip\right\rangle $ and $\left\langle \chipp\right\rangle $ of a Kramers polycrystalline sample in the applied dc field: 

\begin{gather}
\left\langle \chi'\right\rangle =\intop_{0}^{1}dx\frac{x^{2}\left[h_{dc}^{2}e^{-h_{dc}^{2}x^{2}/2}\left(1-x^{2}\right)+\alpha^{2}h_{dc}^{4}x^{4}\right]^{2}}{\left[h_{dc}^{2}e^{-h_{dc}^{2}x^{2}/2}\left(1-x^{2}\right)+\alpha^{2}h_{dc}^{4}x^{4}\right]^{2}+w^{2}},\label{eq:in-phase susceptibility - Kramers powder - no dipolar field}
\end{gather}

\begin{gather}
\left\langle \chi''\right\rangle =\intop_{0}^{1}dx\frac{x^{2}w\left[h_{dc}^{2}e^{-h_{dc}^{2}x^{2}/2}\left(1-x^{2}\right)+\alpha^{2}h_{dc}^{4}x^{4}\right]}{\left[h_{dc}^{2}e^{-h_{dc}^{2}x^{2}/2}\left(1-x^{2}\right)+\alpha^{2}h_{dc}^{4}x^{4}\right]^{2}+w^{2}}.\label{eq:out-of-phase susceptibility - Kramers powder- no dipolar field}
\end{gather}

\begin{figure}
\includegraphics{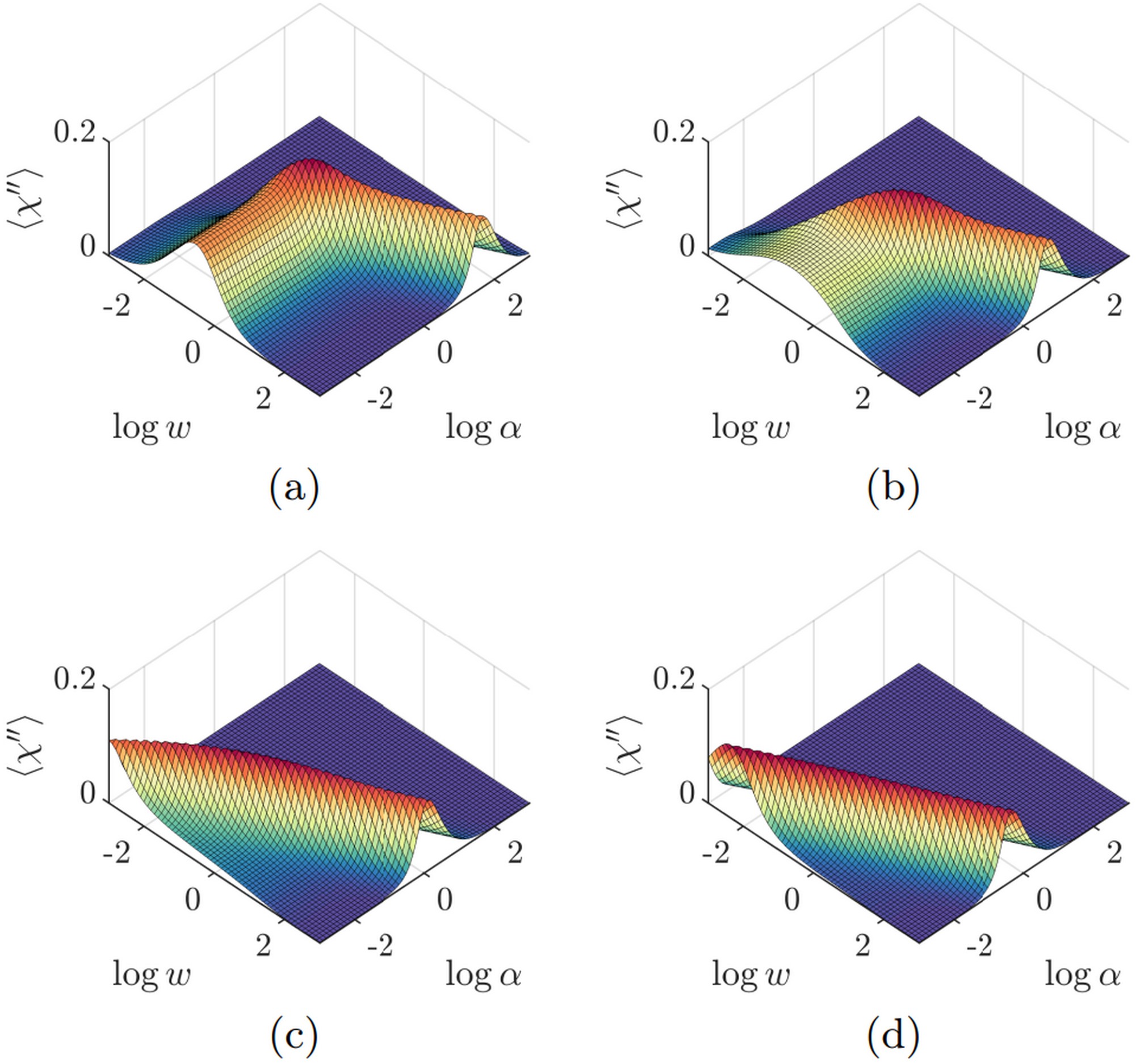}

\caption{Out-of-phase susceptibility of a Kramers polycrystalline sample sample ($\chi_{0}$ units) without an internal dipolar field but with an applied dc field $h_{dc}=1$ (a), 3 (b), 6 (c), and 10 (d).\label{fig:chipp - Kramer - no dipolar field}}
\end{figure}

In order to find out whether two maxima can arise in $\left\langle \chipp\right\rangle $, we examine it over broad domains of parameters as before. In Fig. \ref{fig:chipp - Kramer - no dipolar field}, we show the variation of $\left\langle \chi''\right\rangle $ versus $\left(\alpha,w\right)$ for several values of $h_{dc}$ in the Kramers polycrystalline samples. The corresponding Cole-Cole plot are shown in the Supplemental Material\cite{Note1}. Similar to the case of non-Kramers polycrystalline samples, a distribution in the orientation of the applied dc field w.r.t. to the anisotropy axis of the molecule cannot give rise to two maxima in $\left\langle \chi''\right\rangle $. As previously, only at $h_{dc}=6,$ a small shoulder is observed. 

In combination with the previous results for the non-Kramers polycrystalline samples, we can conclude that the appearance of two maxima in $\chipp\left(\omega\right)$ cannot come from the sole effect of the spherical uniform distribution of the applied dc field w.r.t. the main anisotropy axis of the molecules. Thus the ``powder EPR'' effect does not show up in the out-of-phase susceptibility signal. This forced us to look for additional ingredients to our model. The first in the list of unaccounted interactions is the internal dipolar field, whose effect will be investigated in the next section.

\section{Low-temperature dynamic susceptibility in the presence of internal dipolar field}

Taking into account the distribution of the dipolar mean field $\vt H_{d}$, the in-phase $\chi'$ and out-of-phase susceptibility $\chi''$ in Eqs. \eqref{eq:crystal in-phase susceptibility - no internal field} and \eqref{eq:crystal out-of-phase susceptibility - no internal field}, respectively, become
\begin{gather}
\chi'=4\int d^{3}H_{d}\,p\left(H_{d,x}\right)p\left(H_{d,y}\right)p\left(H_{d,z}\right)\nonumber \\
\qquad\qquad\qquad\qquad\qquad\qquad\times\frac{c\Gamma_{mm'}^{2}\cos^{2}\theta}{\left(1+c\right)^{2}\Gamma_{mm'}^{2}+\omega^{2}},\label{eq:crystal in-phase susceptibility - general case}\\
\chi''=4\int d^{3}H_{d}\,p\left(H_{d,x}\right)p\left(H_{d,y}\right)p\left(H_{d,z}\right)\nonumber \\
\qquad\qquad\qquad\qquad\quad\times\frac{c}{c+1}\frac{\omega\Gamma_{mm'}\cos^{2}\theta}{\left(1+c\right)^{2}\Gamma_{mm'}^{2}+\omega^{2}}.\label{eq:crystal out-of-phase susceptibility - general case}
\end{gather}

\subsection{Non-Kramers systems}

Substituting the expressions of $p\left(H_{d,\alpha}\right)$, Eq. \eqref{eq:p(H_d,a)}, and $\Gamma_{mm'}$ of non-Kramers system, Eqs. (\ref{eq:Gamma_mm'-1}-\ref{eq:Gamma_mm'-3}), into Eqs. \eqref{eq:crystal in-phase susceptibility - general case} and \eqref{eq:crystal out-of-phase susceptibility - general case}, then defining two additional dimensionless quantities: 
\begin{gather}
h_{dm}\equiv H_{dm}/H_{n},\,h_{z}=h_{dc}x+h_{d,z},
\end{gather}
we obtain the ac susceptibilities for a single-crystal sample:  
\begin{gather}
\chi'=x^{2}\intop_{-\infty}^{+\infty}dh_{z}\frac{e^{-\left(h_{z}-h_{dc}x\right)^{2}/2h_{dm}^{2}}}{\sqrt{2\pi}h_{dm}}\frac{\left(e^{-h_{z}^{2}/2}+\alpha^{2}h_{z}^{2}\right)^{2}}{\left(e^{-h_{z}^{2}/2}+\alpha^{2}h_{z}^{2}\right)^{2}+w^{2}},\label{eq:crystal in-phase susceptibility - Non-Kramers}\\
\chi''=x^{2}\intop_{-\infty}^{+\infty}dh_{z}\frac{e^{-\left(h_{z}-h_{dc}x\right)^{2}/2h_{dm}^{2}}}{\sqrt{2\pi}h_{dm}}\frac{w\left(e^{-h_{z}^{2}/2}+\alpha^{2}h_{z}^{2}\right)}{\left(e^{-h_{z}^{2}/2}+\alpha^{2}h_{z}^{2}\right)^{2}+w^{2}},\label{eq:crystal out-of-phase susceptibility - non-Kramers}
\end{gather}

The averaging of the above expressions for different molecular orientations in polycrystalline samples are trivial by using Eqs.\eqref{eq:powder in-phase susceptibility - general case} and \eqref{eq:powder out-of-phase susceptibility - general case}. As before, the appearance/disappearance of two maxima in $\chipp\left(\omega\right)$ and $\left\langle \chipp\left(\omega\right)\right\rangle $ are studied by considering examples with typical values of the parameters. The case of zero and non-zero applied dc field will be examined separately.

\subsubsection{Zero applied dc field}

In zero applied dc field, the in-phase and out-of-phase susceptibility of the single-crystal and polycrystalline (powder) sample, Eqs. can be straightforwardly rewritten from Eqs. \eqref{eq:crystal in-phase susceptibility - Non-Kramers} and \eqref{eq:crystal out-of-phase susceptibility - non-Kramers}: 
\begin{gather}
\chi'=2I_{\chip}x^{2},\chi''=2I_{\chipp}x^{2}\nonumber \\
\left\langle \chi'\right\rangle =\frac{2}{3}I_{\chip},\left\langle \chi''\right\rangle =\frac{2}{3}I_{\chipp}.\label{eq:polycrystalline non-Kramers zero external field}\\
I_{\chip}\equiv\intop_{0}^{+\infty}dh_{z}\frac{e^{-h_{z}^{2}/2h_{dm}^{2}}}{\sqrt{2\pi}h_{dm}}\frac{\left(e^{-h_{z}^{2}/2}+\alpha^{2}h_{z}^{2}\right)^{2}}{\left(e^{-h_{z}^{2}/2}+\alpha^{2}h_{z}^{2}\right)^{2}+w^{2}},\label{eq:in-phase susceptibility integral}\\
I_{\chipp}\equiv\intop_{0}^{+\infty}dh_{z}\frac{e^{-h_{z}^{2}/2h_{dm}^{2}}}{\sqrt{2\pi}h_{dm}}\frac{w\left(e^{-h_{z}^{2}/2}+\alpha^{2}h_{z}^{2}\right)}{\left(e^{-h_{z}^{2}/2}+\alpha^{2}h_{z}^{2}\right)^{2}+w^{2}},\label{eq:out-of-phase susceptibility integral}
\end{gather}

As in Sec. \ref{sec:No dipolar field}, we vary the model parameters $\alpha$ and $w$  in the same domains  as before. Meanwhile, being the ratio between the dipolar field and hyperfine field, $h_{dm}$ typically runs from 0.1 to 10 and is subject to the effect of dilution \cite{Gatteschi2006}.

\begin{figure}
\includegraphics{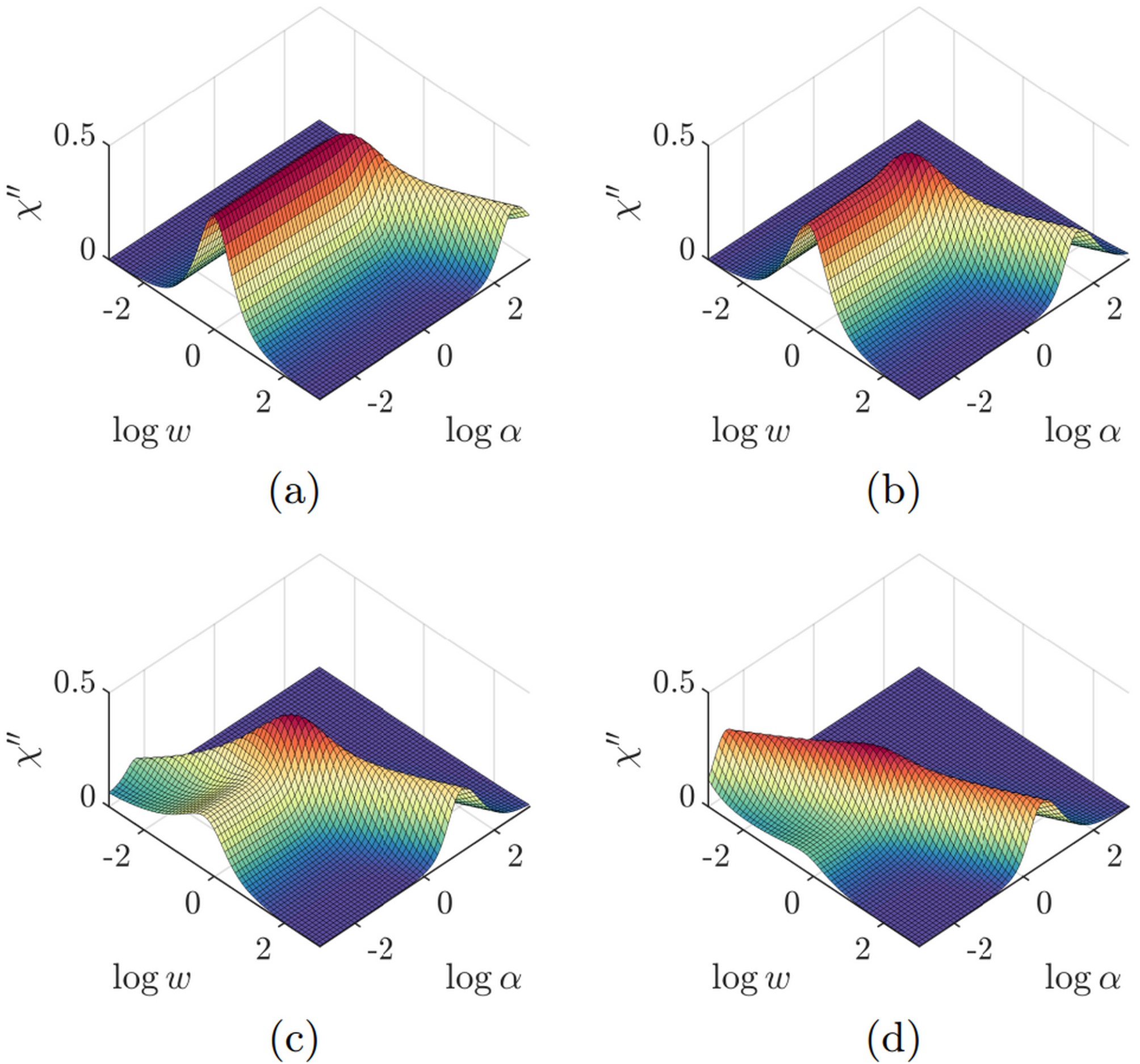}

\caption{Out-of-phase susceptibility of a non-Kramers single-crystal sample in $x^{2}\chi_{0}$ unit with $h_{dc}=0$ and $h_{dm}=0.1$ (a), 1 (b), 3 (c), and 10 (d). \label{fig:chipp - non-Kramer - hdc=00003D0}}
\end{figure}

\emph{Single-crystal samples} - Fig. \ref{fig:chipp - non-Kramer - hdc=00003D0} shows the variation of $\chipp$ vs. $\left(\alpha,w\right)$ in the logarithmic scale for several values of $h_{dm}$ in the single crystal. The corresponding in-phase susceptibilities and Cole-Cole plots are shown in the Supplemental Material. From this figure, it can be seen that a small standard deviation of the dipolar field distribution $h_{dm}=0.1$ or 1 cannot give rise to a second maxima in $\chipp$. This is expected since the smaller $h_{dm}$, the narrower the Gaussian distribution of the dipolar field. Accordingly, for small $\alpha$, the quantum tunneling process, which maximizes at $h_{z}=0$, dominates the relaxation over nearly the whole meaningful 3-sigma domain of the distribution. As a result, only one maximum in $\chipp\left(w\right)$ can appears at $w\sim\mathcal{O}\left(1\right)$. In contrast, for a large $\alpha$, the direct process contribution to $\chipp\left(w\right)$ characterized by $\alpha^{2}h_{z}^{2}$ prevails over the quantum tunneling process in the meaningful domain of the distribution, even with a small $h_{dm}$, which then causes the maximum's frequency location shifted toward a larger $w$ as shown in all subfigures of Fig. \ref{fig:chipp - non-Kramer - hdc=00003D0}.

As the dipolar field distribution widened with a larger $h_{dm}$, the effect of the direct process in the domain of small $w$ and $\alpha$ is getting more apparent. This is demonstrated with the appearance of a second maximum in $\chipp$ as shown in Fig. \ref{fig:chipp - non-Kramer - hdc=00003D0}c and \ref{fig:chipp - non-Kramer - hdc=00003D0}d. While the maximum of $\chipp\left(w\right)$ at $w\sim\mathcal{O}\left(1\right)$ can be explained by the dominance of the quantum tunneling process in the vicinity of the peak of the Gaussian distribution, the second maximum at the smaller frequency appears due to the expansion of the distribution to the domain where $h$ is sufficiently large to make $e^{-h_{z}^{2}/2}$ negligible comparing to $\alpha^{2}h_{z}^{2}$ but still small enough for the Gaussian probability noticeable. However, this does not mean that the wider the dipolar field distribution, the more likely the two maxima emerge. In fact, there exists some limiting value of $h_{dm}$ where a broad distribution will weaken the role of the quantum tunneling process in $\chipp$ since the quantum tunneling process contribution to the integral of $\chipp$ is nearly unchanged due to $e^{-h_{z}^{2}/2}$ exponentially decreased, whereas the contribution of the direct process to $\chipp$ becomes larger due to a larger corresponding integration domain. Consequently, $\chipp$ at the high-frequency maximum relatively decreases comparing to low-frequency maximum. This effect can be observed from Fig. \ref{fig:chipp - non-Kramer - hdc=00003D0}c and \ref{fig:chipp - non-Kramer - hdc=00003D0}d where the high-frequency maximum in the case $h_{dm}=10$ becomes less clear than in the case $h_{dm}=3$. Consequently, there only exists a \emph{limited} favorable range of the parameters $\alpha$ and $h_{dm}$ for detecting two maxima in $\chipp\left(w\right)$ at zero applied dc field.

\emph{Polycrystalline (powder) samples} - From Eq. \eqref{eq:polycrystalline non-Kramers zero external field}, it is obvious that the dynamic susceptibility expressions of the polycrystalline samples only differs from the single crystal's by just a constant factor. Hence, the same behaviors of the dynamic susceptibility are expected. 

\subsubsection{Non-zero applied dc field}

\begin{figure}
\includegraphics{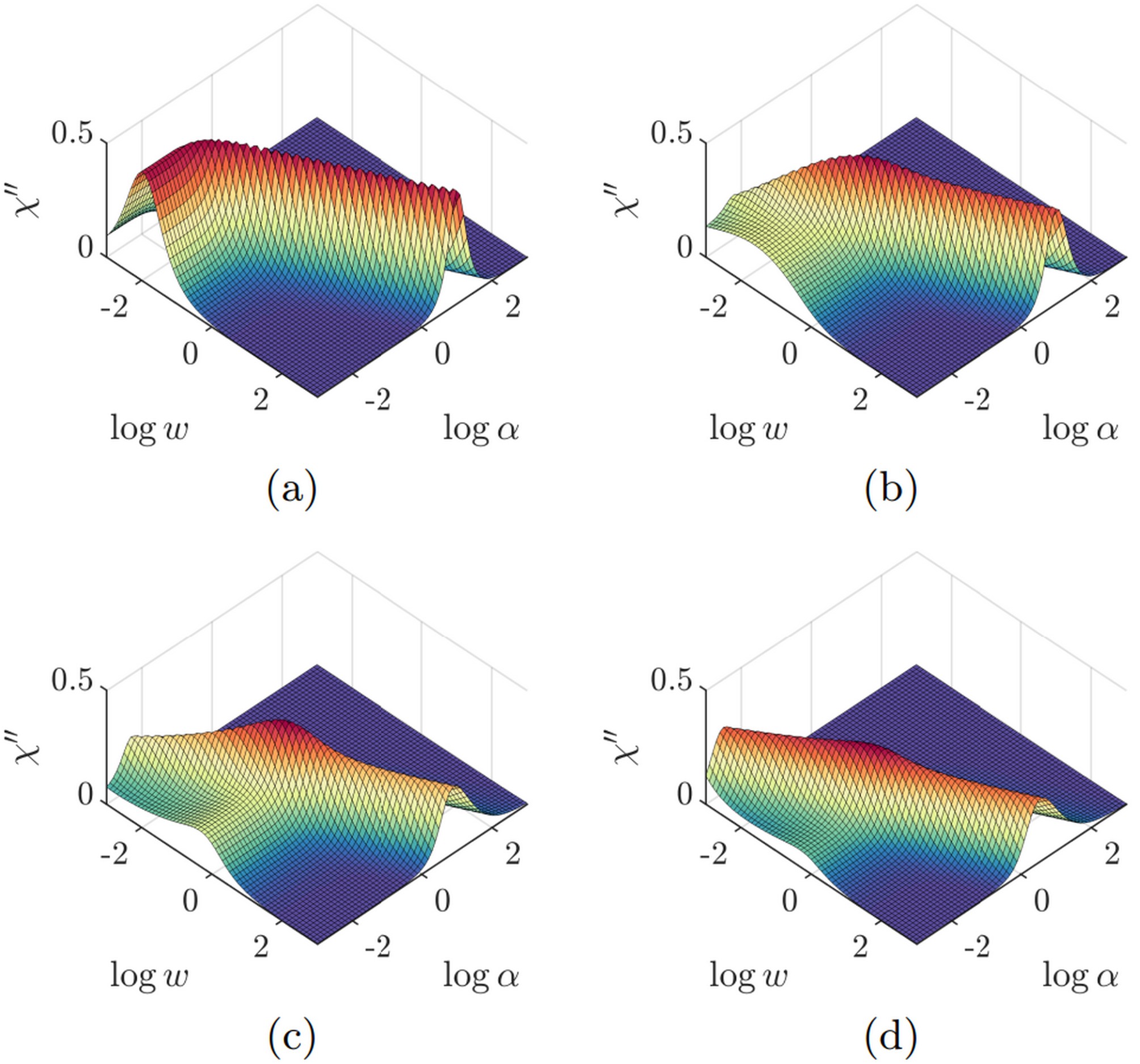}

\caption{Out-of-phase susceptibility of a non-Kramers single-crystal sample in $x^{2}\chi_{0}$ unit with $h_{dc}\cos\theta=3$ and $h_{dm}=0.1$ (a), 1 (b), 3 (c), and 10 (d) .\label{fig:chipp - non-Kramer - hdcx=00003D3}}
\end{figure}

\emph{Single-crystal samples} - $\chip$ and $\chipp$ of a single-crystal sample in this case have the form of Eqs. \eqref{eq:crystal in-phase susceptibility - Non-Kramers} and \eqref{eq:crystal out-of-phase susceptibility - non-Kramers}, respectively. The applied dc field effectively causes a shift in the expectation value of the Gaussian distribution in these expressions from zero value. This reduces the contribution from the quantum tunneling process to $\chipp$ while concomitantly increases the one from the direct process. Accordingly, with a small $h_{dc}x$, an appearance of two maxima in $\chipp$ at lower values of $h_{dm}$  comparing to the previous case is expected. This behavior can be seen from Fig. \ref{fig:chipp - non-Kramer - hdcx=00003D3}b showing $\chipp$ at $h_{dc}x=3$ where the second maximum appears at a lower $h_{dm}=1$ comparing to $h_{dm}=3$ in the zero applied dc field. Apparently, in the case of a large $h_{dc}x$, the contribution from the direct process may dominate over the one from the quantum tunneling and only one maximum of $\chipp$ may be obtained unless the dipolar field distribution width is wide enough. These can be seen from Fig. S5 and S7 in the Supplemental Material \cite{Note1}\emph{ }which\emph{ }correspond to $h_{dc}x=6$ and 10. 

It is also worth noting that with a fixed $h_{dc}x$, increasing $h_{dm}$ produces the same behavior with the same reason as in the previous case of zero external field. Particularly, at first the broadening of the distribution facilitates the appearance of two maxima but then a large $h_{dm}$ would lead to a reduction in the magnitude of the high-frequency maximum associated with the quantum tunneling contribution. 

From the figures of both zero and non-zero applied dc field, it is noticeable that as two maxima exist, the high-frequency maximum location $w_{\mathrm{max,hf}}$ is insensitive to the value of the parameter $\alpha$. Moreover, it is often the case that $w_{\mathrm{max,hf}}\sim\mathcal{O}\left(1\right)$ (or equivalently $\omega_{\mathrm{max,hf}}\sim\mathcal{O}\left(2\Gamma_{tn,0}\right)$) for both crystal or polycrystalline (powder) samples. This results from the intuitive fact that, for two prospective maxima to coexist, they must be far away to not merge into one. Accordingly, there must have been two domains of frequency where one relaxation process suppresses the other. Eq. \eqref{eq:crystal out-of-phase susceptibility - non-Kramers} then can be roughly approximated as: 
\begin{multline}
\chi''/x^{2}\approx\intop_{-h_{1}}^{h_{1}}dh_{z}\frac{e^{-\left(h_{z}-h_{dc}x\right)^{2}/2h_{dm}^{2}}}{\sqrt{2\pi}h_{dm}}\frac{we^{-h_{z}^{2}/2}}{e^{-h_{z}^{2}}+w^{2}}\\
+\left(\intop_{-\infty}^{-h_{2}}+\intop_{h_{2}}^{+\infty}\right)dh_{z}\frac{e^{-\left(h_{z}-h_{dc}x\right)^{2}/2h_{dm}^{2}}}{\sqrt{2\pi}h_{dm}}\frac{\left(w/\alpha^{2}\right)h_{z}^{2}}{h_{z}^{4}+\left(w/\alpha^{2}\right)^{2}},\label{eq:approximation explains observed behaviors}
\end{multline}
where $h_{1}$ and $h_{2}$ are two positive limits satisfying $\exp\left(-h_{1}^{2}/2\right)\gg\alpha^{2}h_{1}^{2}$ and $\exp\left(-h_{2}^{2}/2\right)\ll\alpha^{2}h_{2}^{2}$ respectively. This approximation results from the fact that the term $e^{-h_{z}^{2}/2}$ ($\alpha^{2}h_{z}^{2}$) fast decreases (increases), which then leads to a narrow $\left[-h_{2},-h_{1}\right]\cup\left[h_{1},h_{2}\right]$ domain and accordingly a negligible value of the integral in the respective domain. Conforming with the previous analysis, we associate the first integral with the high-frequency maximum and the second with the low-frequency maximum. An application of the mean value theorem for integrals to the first integral yields 
\begin{gather}
I_{1}=\frac{we^{-h_{m}^{2}/2}}{e^{-h_{m}^{2}}+w^{2}}\intop_{-h_{1}}^{h_{1}}dh_{z}\frac{e^{-\left(h_{z}-h_{dc}x\right)^{2}/2h_{dm}^{2}}}{\sqrt{2\pi}h_{dm}},
\end{gather}
where $h_{m}\in\left(-h_{1},h_{1}\right)$. Since the term $e^{-h_{z}^{2}/2}$ ($\alpha^{2}h_{z}^{2}$) decreases (increases) fast w.r.t. $h_{z}$, $h_{1}$ and $h_{2}$ in effect are of $\mathcal{O}\left(1\right)$ for our investigated $\alpha$ domain, i.e. insensitive to the value of $\alpha$. Hence, a variation of $\alpha$ will hardly change the high-frequency maximum location.  In order to know the order of magnitude of the frequency of the first maximum, since $h_{1}\sim\mathcal{O}\left(1\right)$, we can further take the middle point $h_{z}=0$ of the domain $\left[-h_{1},h_{1}\right]$ as $h_{m}$ for a rough approximation. This thus yields $w_{\mathrm{max,hf}}\approx\mathcal{O}\left(1\right)$ as observed. 

Furthermore, from the second integral of Eq. \eqref{eq:approximation explains observed behaviors}, further denoted $I_{2}$, we can also infer the rough linearity of the low-frequency maximum's location $w_{\mathrm{max,lf}}$ on the $\log w$ vs. $\log\alpha$ diagram. In fact, since the location of the maximum of $I_{2}\left(w'\equiv w/\alpha^{2}\right)$, $w'_{max}$, weakly depends on $\alpha$ (via $h_{2}$), then $w_{\mathrm{max,lf}}\sim\alpha^{2}$ or $\log w_{\mathrm{max,lf}}\sim\log\alpha$ as observed in the figures. Similarly, the same approximation can also be used for the explanation of the the linearity of $\log w_{\mathrm{max}}$ vs. $\log\alpha$ in the large $\alpha$-domain where only one maximum  exists.

\emph{Polycrystalline (powder) samples} - For this kind of sample, the effect of relative orientation between the magnetic field and the anisotropy main axis is averaged out. Consequently, with the same values of $h_{dc}$ as before for $h_{dc}x$, the effect of the applied dc field on reducing (decreasing) the contribution of the quantum tunneling process (the direct process) is alleviated. This can be seen from comparing the out-of-phase susceptibility plots of polycrystalline sample included in the Supplementary Material with the ones from the single-crystal. Besides this effect, in general, the behavior of the dynamic susceptibility under parameters variations, and accordingly the appearance/disappearance of two maxima in $\chipp$, of the polycrystalline sample is not different from the single-crystal one.

\subsection{Kramers systems}

Substituting the expression of $\Gamma_{mm'}$ for the Kramers system, Eq. \eqref{eq:total relaxation rate - Kramers system}, into Eqs. \eqref{eq:crystal in-phase susceptibility - general case} and \eqref{eq:crystal out-of-phase susceptibility - general case} results in the expressions of $\chip$ and $\chipp$ for Kramers single-crystal samples:

\begin{multline}
\chip=\frac{\cos^{2}\theta}{\left(\sqrt{2\pi}h_{dm}\right)^{3}}\int dh_{x}dh_{y}dh_{z}\\
\times e^{-\left[\left(h_{x}-h_{dc}\sin\theta\cos\varphi\right)^{2}+\left(h_{y}-h_{dc}\sin\theta\sin\varphi\right)^{2}+\left(h_{z}-h_{dc}\cos\theta\right)^{2}\right]/2h_{dm}^{2}}\\
\times\frac{\left[\left(h_{x}^{2}+h_{y}^{2}\right)e^{-h_{z}^{2}/2}+\alpha^{2}h_{z}^{4}\right]^{2}}{\left[\left(h_{x}^{2}+h_{y}^{2}\right)e^{-h_{z}^{2}/2}+\alpha^{2}h_{z}^{4}\right]^{2}+w^{2}},\label{eq:in-phase susceptibility - Kramers crystal}
\end{multline}
and

\begin{multline}
\chi''=\frac{\cos^{2}\theta}{\left(\sqrt{2\pi}h_{dm}\right)^{3}}\int dh_{x}dh_{y}dh_{z}\\
\times e^{-\left[\left(h_{x}-h_{dc}\sin\theta\cos\varphi\right)^{2}+\left(h_{y}-h_{dc}\sin\theta\sin\varphi\right)^{2}+\left(h_{z}-h_{dz}\cos\theta\right)^{2}\right]/2h_{dm}^{2}}\\
\qquad\qquad\qquad\times\frac{w\left[\left(h_{x}^{2}+h_{y}^{2}\right)e^{-h_{z}^{2}/2}+\alpha^{2}h_{z}^{4}\right]}{\left[\left(h_{x}^{2}+h_{y}^{2}\right)e^{-h_{z}^{2}/2}+\alpha^{2}h_{z}^{4}\right]^{2}+w^{2}}.\label{eq:out-of-phase susceptibility - Kramers crystal}
\end{multline}
where $h_{\alpha}\equiv h_{dc,\alpha}+h_{d,\alpha},\,\alpha=x,y,z$. For polycrystalline samples, we can average over the orientation between the field and the main anisotropy axis of the microcrystals using Eqs. \eqref{eq:powder in-phase susceptibility - general case} and \eqref{eq:powder out-of-phase susceptibility - general case}.

Investigation of the behavior of the dynamic susceptibility for Kramers system is done similarly to the case of non-Kramers systems in the previous section.

\subsubsection{Zero applied dc field}

\emph{Single-crystal sample} - Normally, for a Kramers system and in zero magnetic field, the time-reversal symmetry inhibits the relaxation. However, due to the existence of the internal field, the relaxation still happens in this case. The corresponding dynamic susceptibility can be easily simplified from Eq. \eqref{eq:in-phase susceptibility - Kramers crystal} and \eqref{eq:out-of-phase susceptibility - Kramers crystal}: 
\begin{multline}
\chip=\frac{2\cos^{2}\theta}{\sqrt{2\pi}h_{dm}^{3}}\intop_{0}^{+\infty}\intop_{0}^{+\infty}dh_{\perp}dh_{z}\,e^{-\left(h_{\perp}^{2}+h_{z}^{2}\right)/2h_{dm}^{2}}\\
\times\frac{h_{\perp}\left(h_{\perp}^{2}e^{-h_{z}^{2}/2}+\alpha^{2}h_{z}^{4}\right)^{2}}{\left(h_{\perp}^{2}e^{-h_{z}^{2}/2}+\alpha^{2}h_{z}^{4}\right)^{2}+w^{2}}\label{eq:in-phase susceptibility - Kramers crystal-hdc=00003D0}
\end{multline}

\begin{multline}
\chi''=\frac{2\cos^{2}\theta}{\sqrt{2\pi}h_{dm}^{3}}\intop_{0}^{+\infty}\intop_{0}^{+\infty}dh_{\perp}dh_{z}\,e^{-\left(h_{\perp}^{2}+h_{z}^{2}\right)/2h_{dm}^{2}}\\
\qquad\qquad\qquad\times\frac{wh_{\perp}\left(h_{\perp}^{2}e^{-h_{z}^{2}/2}+\alpha^{2}h_{z}^{4}\right)}{\left(h_{\perp}^{2}e^{-h_{z}^{2}/2}+\alpha^{2}h_{z}^{4}\right)^{2}+w^{2}}.\label{eq:out-of-phase susceptibility - Kramers crystal-hdc=00003D0}
\end{multline}

\begin{figure}
\includegraphics{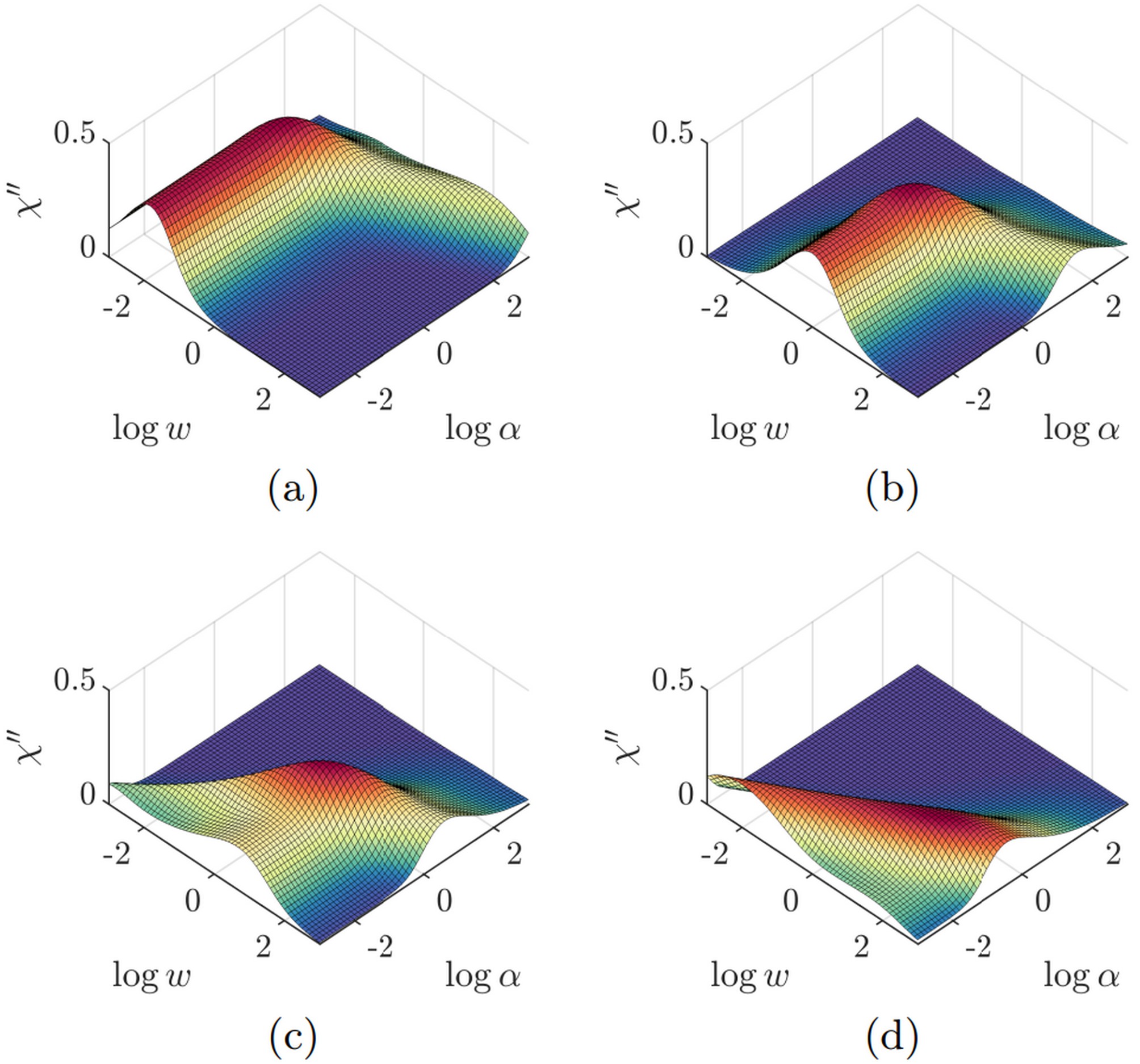}

\caption{Out-of-phase susceptibility of a Kramers single-crystal sample in $x^{2}\chi_{0}$ unit without an external field and $h_{dm}=0.1$ (a), 1 (b), 3 (c), and 10 (d).\label{fig:chipp - Kramer - hdc=00003D0 - crystal}}
\end{figure}

 Fig. \ref{fig:chipp - Kramer - hdc=00003D0 - crystal} shows the variation of $\chipp$ for several values of $h_{dm}$. We can see that the behavior of $\chipp$ is basically the same as for the non-Kramers systems, with a second maximum or a shoulder appearing at large $h_{dm}$. As explained, this comes from the expansion of the dipolar field distribution which accordingly redistributes the contributions to the out-of-phase susceptibility between the direct and quantum tunneling process. The second maximum will appear as the contribution of the quantum tunneling and the direct process to the out-of-phase susceptibility are of the same order of magnitude. 

\emph{Polycrystalline (powder) sample} - in zero applied dc field, it is obvious that there is no difference in the behavior of the dynamic susceptibility under parameters variations between the single-crystal and polycrystalline samples since the expressions of the dynamic susceptibility of these two cases are only different by a constant factor. 

\subsubsection{Non-zero applied dc field}

\emph{Single-crystal sample} - Since the number of parameters defining $\chip$ and $\chipp$ rises to five ($\alpha$, $h_{dm}$, $h_{dc}$, $\theta$, and $\varphi$),  we consider only one direction of the applied magnetic oriented at an angle $\theta=\phi=\pi/4$ for simplicity. As before, the cases with $h_{dc}=3$ (and $h_{dc}=6,10$ in the Supplemental Material) is investigated.

\begin{figure}
\includegraphics{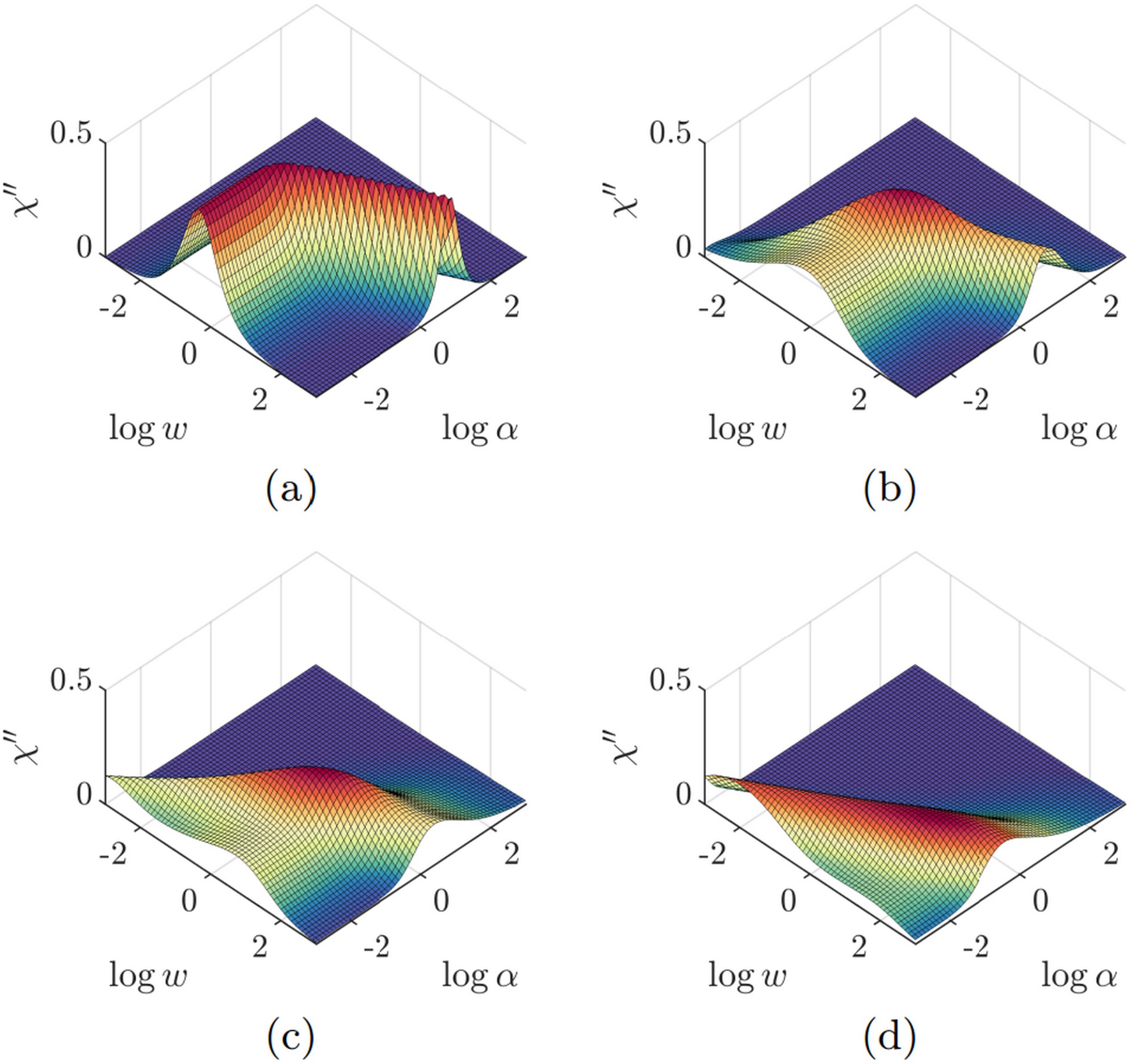}

\caption{Out-of-phase susceptibility of a Kramers single-crystal sample in $\chi_{0}$ unit with $h_{dc}=3$ and $h_{dm}=0.1$ (a), 1 (b), 3 (c), and 10 (d).\label{fig:chipp - Kramer - hdc=00003D3 - crystal}}
\end{figure}

Contrary to non-Kramers systems where an applied dc field increases the direct process transition rate but decrease the quantum tunneling process, in Kramers systems, this applied dc field not only increases the direct process transition rate but may also  increase the quantum tunneling rate as the tunneling gap is proportional to the dc field magnitude. However, since the direct process rate is proportional to a higher power (quadruple) of the field magnitude, it always increases faster. Accordingly, the relative contribution to $\chipp$ of the direct process is proportional to the dc amplitude. This is demonstrated in Fig. \ref{fig:chipp - Kramer - hdc=00003D3 - crystal}b, which shows a slight emergence of a shoulder at low $w$ and low $\alpha$ domain, contrary to the case of $h_{dc}=0$ in Fig. \ref{fig:chipp - Kramer - hdc=00003D0 - crystal}b. Reinforced by an expansion of the dipolar field distribution, this shoulder then transforms into a second maximum in Fig. \ref{fig:chipp - Kramer - hdc=00003D3 - crystal}c ($h_{dc}=h_{dm}=3$) and Fig. \ref{fig:chipp - Kramer - hdc=00003D3 - crystal}d ($h_{dc}=3$, $h_{dm}=10$). From Fig. \ref{fig:chipp - Kramer - hdc=00003D3 - crystal}c, it can also be seen that the relative magnitude of the low-frequency maximum to the high-frequency maximum is increased in comparison to the same case in Fig. \ref{fig:chipp - Kramer - hdc=00003D0 - crystal}c ($h_{dc}=0$, $h_{dm}=3$), which indicates that the direct process contribution to $\chipp$ is strengthened. However, a comparison of Fig. \ref{fig:chipp - Kramer - hdc=00003D0 - crystal}d ($h_{dc}=0$, $h_{dm}=10$) and Fig. \ref{fig:chipp - Kramer - hdc=00003D3 - crystal}d ($h_{dc}=3$, $h_{dm}=10$) shows that the effect of the applied dc field is mitigated with a large distribution width and the relative magnitude between two maxima of $\chipp$ for both cases becomes virtually indistinguishable. This is expected since a large distribution width of the internal field will partially negate the effect of the applied dc field.

It is worth mentioning that with a fixed $h_{dc}$, the behavior of $\chipp$ w.r.t. $h_{dm}$ variation, and accordingly the appearance of two maxima in $\chipp$, is basically the same as in zero dc field. This is obvious considering that the nature of the appearance of two maxima are the same in both cases, which is nothing but the effect of the dipolar field distribution of which different domains favor different relaxation processes. On the other hand, fixing $h_{dm}$ while increasing $h_{dc}$ will increase (suppress) the contribution of the direct process (quantum tunneling) to $\chipp$. Certainly, depending on the relative magnitude between $h_{dc}$ and $h_{dm}$, the behavior of $\chipp$ will vary. For more details, a broader set of figures corresponding to the cases $h_{dc}=6,10$ are provided in the Supplemental Material.

Similarly to the non-Kramers systems, in the case when two maxima exist (for both zero and non-zero applied dc field), the location of the high-frequency maximum varies little w.r.t. the change of the parameters $h_{dc}$, $h_{dm}$, and $\alpha$, while the low-frequency maximum's location is proportional to $\alpha$. This behavior can be also explained by a similar approximation to Eq. \eqref{eq:approximation explains observed behaviors}. A minor difference here is that instead of an integration over the domains $\left(-h_{1},h_{1}\right)$ and $\left(-\infty,-h_{2}\right)\cup\left(h_{2},+\infty\right)$, the integration in the Eq. \eqref{eq:out-of-phase susceptibility - Kramers crystal} is approximated over two \emph{volume} domains where $\left(h_{x}^{2}+h_{y}^{2}\right)e^{-h_{z}^{2}/2}\gg\alpha^{2}h_{z}^{4}$ and $\left(h_{x}^{2}+h_{y}^{2}\right)e^{-h_{z}^{2}/2}\ll\alpha^{2}h_{z}^{4}$.

\emph{Polycrystalline (powder) sample} - An averaging over the relative orientation of the magnetic field and the crystal main anisotropy axis eliminates the $\left(\theta,\varphi\right)$ dependence of dynamic susceptibility. Besides trivially reducing the absolute value of $\chipp$, another difference to the single-crystal case with $\theta=\varphi=\pi/4$ is that the effect of the applied dc field to increasing the direct process contribution to $\chipp$ seems stronger. This is showcased at virtually any $h_{dc}$. For examples, in Fig. S21b at $\left(h_{dc},h_{dm}\right)=\left(3,1\right)$, the shoulder at low $w$ and low $\alpha$ corresponding to the low-frequency maximum becomes more pronounced comparing to the corresponding Fig. \ref{fig:chipp - Kramer - hdc=00003D3 - crystal}b of the single-crystal case; or at $\left(h_{dc},h_{dm}\right)=\left(6,3\right)$ (Fig. S23c), the relative magnitude of the high-frequency maximum compared to the low-frequency maximum becomes weaker than in the case of single-crystal samples (see Supplemental Material \cite{Note1}).  However, the dependence of $\chipp$ on $h_{dc}$ and $h_{dm}$ in general is similar to the case of single-crystal samples, which indicates no difference in the mechanism of the phenomenon in two kinds of samples.

\section{Multiple maxima in $\protect\chipp$ and its correlation with applied field, temperature, and magnetic dilution \label{sec:Theory vs. experiment}}

As clear from the model, the necessary condition for the appearance of multiple maxima in the dynamic susceptibility of both non-Kramers and Kramers SMM samples is the existence of a sufficiently wide dipolar field distribution which has two domains where one relaxation process effectively dominates the other(s). Since a variation of the applied dc field, sample dilution, or temperature will alter the dipolar field distribution and/or components of the total relaxation rate, it is apparent that these will significantly affect to the observation of the multiple maxima in the dynamic susceptibility. In this section, hence, we will investigate these experimental factors within the proposed mechanism.  

Before moving on, it should be noted that in the previous sections, for simplicity we have demonstrated the mechanism by supposing that the former relaxation process is the direct process and the latter is the quantum tunneling. However, the mechanism is not restricted to this assumption. From the nature of the mechanism and the formalism of the demonstration, as mentioned in Sec. \ref{sec:Microscopic-description}, it is obvious that the mechanism is also valid if the former one is the Raman process/Orbach process and the latter is the quantum tunneling process as well.

\begin{figure}
\includegraphics{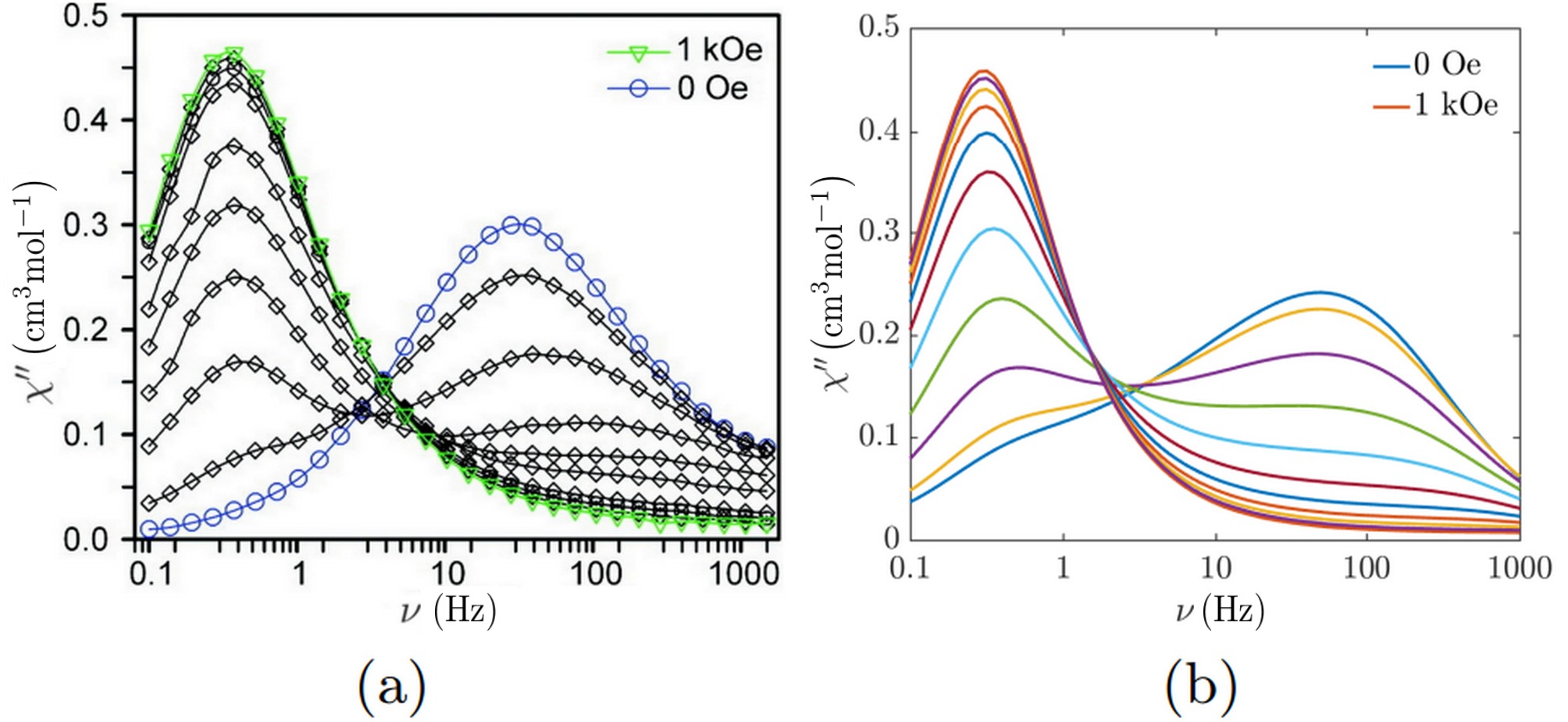}

\caption{(a) Out-of-phase susceptibility of $\mathrm{\left(Ph_{4}P\right)_{2}\left[Co\left(SPh\right)_{4}\right]}$ at $T=2$ K under applied dc field from 0 to 1000 Oe in 100 Oe increments (reprinted with permission from \citet{Zadrozny2011a}, copyright 2011 American Chemical Society); (b) The same quantity generated from our model using an averaging of Eq. \ref{eq:crystal out-of-phase susceptibility - no internal field} for Kramer powder sample, an addition of the second-order Raman process rate $\Gamma_{\mathrm{Raman}}=C_{\mathrm{Raman}}T^{9}$ into the expression of $\Gamma_{mm'}$, Eq. \eqref{eq:Gamma_mm'-1}, and parameters $\alpha=0$, $C_{\mathrm{Raman}}=0.00185\,\mathrm{s^{-1}K^{-9}}$, $H_{n}=75\,\mathrm{Oe}$, $H_{dm}=150\,\mathrm{Oe}$, $\Gamma_{\mathrm{tn},0}=63\,\mathrm{s^{-1}}$, and $\chi_{0}=2.9\,\mathrm{cm^{3}mol^{-1}}$. \label{fig:Zadrozny2011}}
\end{figure}

\emph{Applied dc field} - Consider first the applied field, obviously, increasing the applied dc field $H_{dc}$ is synonymous with increasing $h_{dc}$ in our model. This apparently results in an increase of the relative contribution of the direct process (or Raman process/Orbach process if included) to $\chipp$ comparing to the quantum tunneling process's contribution. At the beginning, this may give rise to two maxima in $\chipp$ but then reverse the situation when the direct process (Raman process/Orbach process) contribution overtakes the one of the quantum tunneling, leaving only one low-frequency maximum in $\chipp$. This behavior is illustrated in Fig. \eqref{fig:Zadrozny2011} where a comparison of the out-of-phase susceptibility $\chipp$ derived from our model with the experimental data from \citet{Zadrozny2011a} is given. Here, we have used the Raman process and the quantum tunneling as two relaxation processes involve in the formation of $\chipp$ while neglected the direct and Orbach process. This comes from the fact that since from the experiment, the slow maximum positions are almost unchanged with the applied field variation (see Fig. \ref{fig:Zadrozny2011}a), it is supposed the direct process is negligible and the slow maximum is caused by the Raman process. Further, since extensive EPR analysis and dc magnetic susceptibility measurements \cite{Fukui1991,Fukui1992,Fukui1995} as well as theoretical computation \cite{Maganas2011} all show that the investigated compound possesses a large negative value $D$ of $70\,\mathrm{cm^{-1}}$, we also exclude the effect of the Orbach relaxation process. As can be seen from Fig. \ref{fig:Zadrozny2011}, our model gives a good description of the appearance/disappearance of the multiple maxima in $\chipp$ of the studied compound. Further, the constant $C_{\mathrm{Raman}}$, the characteristic collective nuclear magnetic field $H_{n}$, and the standard deviation of the dipolar field $H_{dm}$ are also in the typical range of these quantities \cite{Abragam1970,Gatteschi2006} ($C_{\mathrm{Raman}}\in\left[10^{-5},10^{-1}\right]$, $H_{n}$ and $H_{dm}$ are from several tens to hundreds of Oe). The slight difference between experiment and theory can be possibly attributed to the effect of the direct process and the mean field approximation used in the theory, whose role increases with the applied field. Interestingly, this kind of behavior suggested by our model also agrees with other observations from \citet{Lucaccini2016b} (see Fig. 4 therein), \citet{Jeletic2011} (see Fig. 3 therein), or \citet{Li2016} (see Fig. S3 therein).

\emph{Temperature variation} - Another common factor which may induce two maxima in $\chipp$ is the variation of the temperature. Considering that the effects of the temperature on the characteristic collective nuclear field $H_{n}$, and accordingly the quantum tunneling rate, is negligible, temperature variation mainly affects the direct process rate or the Raman process if present (another effect is the increasing involvement of the Orbach process if the first excited (quasi-) doublet is not well separated from the ground (quasi-) doublet).  At zero or weak applied dc field, increasing (reducing) $T$ will then increase (reduce) the relative contribution of the direct (Raman, Orbach) relaxation process(es) due to the enlargement of the magnetic field distribution domain where the relaxation process(es) dominates the quantum tunneling relaxation. Accordingly, at the beginning, this facilitates (hinders) the formation of another slow maximum in $\chipp$ (if not existed yet). After forming two maxima in $\chipp$, keep increasing $T$ will lead to a decreasing distance between two maxima considering that this distance is characterized for the difference in the relaxation rate between the slow and fast relaxation process(es) and this difference is smaller due to the $T$-dependence of the direct (Raman) process and the $T$-independence of the quantum tunneling relaxation. However, at some limit when the contribution from slow relaxation process(es) is dominant, or the distance between two maxima is not sufficiently large to be well separated, the fast one will disappear, leaving only one maximum in $\chipp$. In short, observation of this behavior in $\chipp$ under an increasing temperature will proceed from one fast maximum (with unchanged frequency location) to two maxima (with closer and closer distance between them) then to one slow maximum. It is also worthy to note that temperature variation is not the golden key to make two maxima occur, other conditions, such as the distribution is sufficiently wide, two maximum are of the same order of magnitude or separated far enough but not beyond the instrument's resolution, should be satisfied as well.

\begin{figure}
\includegraphics{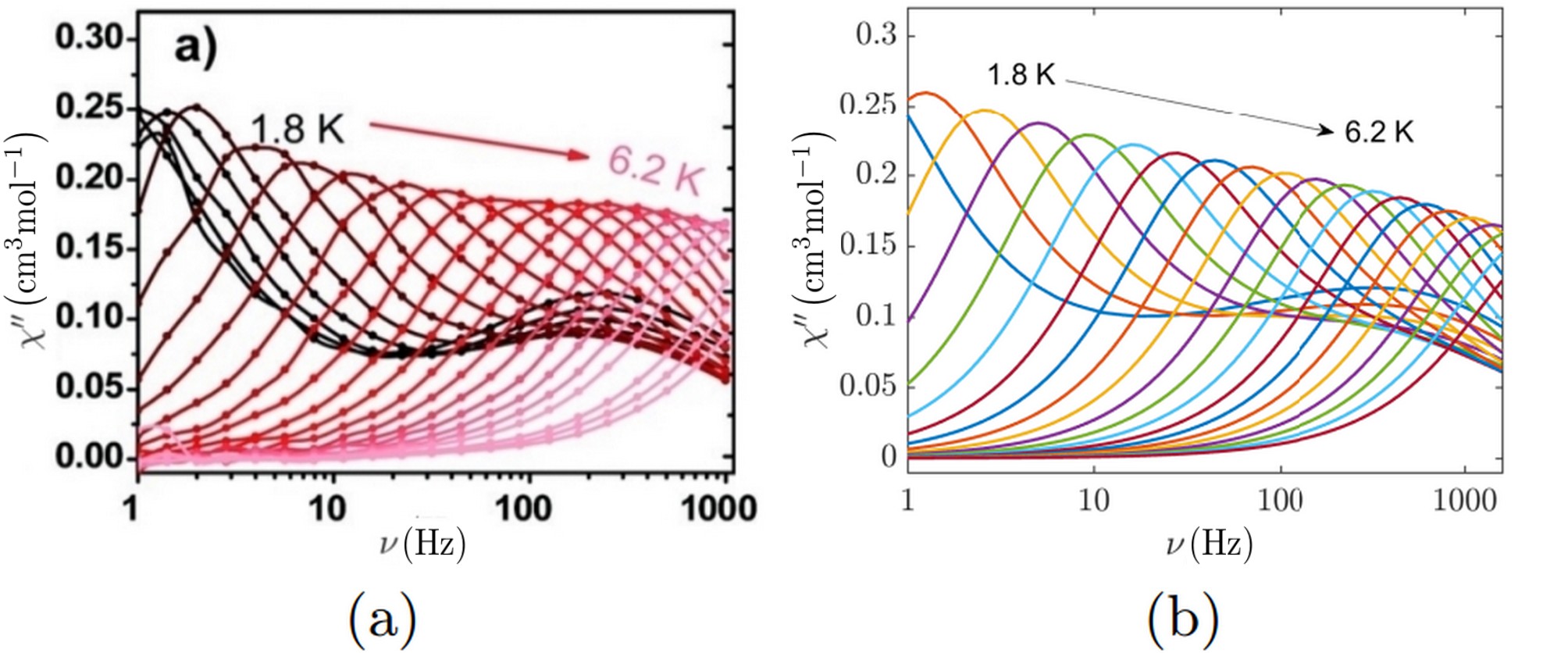}\caption{(a) Frequency dependence of the out-of-phase susceptibility of $\mathrm{\left[Co(Tp^{*})_{2}\right]}$ under 400 Oe applied dc fields for various temperatures from \citet{Li2016}, reproduced by permission of The Royal Society of Chemistry; (b) The same quantity generated from our model using an averaging of Eq. \ref{eq:crystal out-of-phase susceptibility - no internal field} for Kramer powder sample, an addition of the second-order Raman process rate $\Gamma_{\mathrm{Raman}}=C_{\mathrm{Raman}}T^{7}$ into the expression of $\Gamma_{mm'}$, Eq. \eqref{eq:Gamma_mm'-1}, and parameters $\alpha=0$, $C_{\mathrm{Raman}}=0.025\,\mathrm{s^{-1}K^{-7}}$, $H_{n}=125\,\mathrm{Oe}$, $H_{dm}=600\,\mathrm{Oe}$, $\Gamma_{\mathrm{tn},0}=100\,\mathrm{s^{-1}}$, and $\chi_{0}=5.7\,\mathrm{cm^{3}mol^{-1}}$. \label{fig:Li2016}}
\end{figure}

To exemplify the effect of temperature variation in inducing appearance/disappearance of two maxima in $\chipp$, Fig. \ref{fig:Li2016} shows a comparison of the present theory with recent experimental data from \citet{Li2016}. Using a first-order Raman process and ignoring the direct process, which follows from the analysis of the original authors \cite{Li2016}, and a quantum tunneling process, as can be seen from the figure, the theoretical result gives an excellent agreement with the experiment. Value of $C_{\mathrm{Raman}}=0.025\mathrm{\,s^{-1}K^{-7}}$ is also of the same order of magnitude with the fitting from the original work $\left(C_{\mathrm{Raman}}=0.051\mathrm{\,s^{-1}K^{-7.04}}\right)$. The difference of $C_{\mathrm{Raman}}$ probably comes from two sides, firstly from the fitting of the Cole-Cole plots in the experimental work with the \emph{empirical} generalized Debye function \cite{Cole1941,Li2016}, and second from our omission of the direct process to avoid the over-parameterization problem. Other parameters extracted from our simulation are also in their typical range. It should also be noted that according to the \emph{ab initio} calculation\cite{Li2016}, the first excited doublet of the studied compound $\mathrm{\left[Co(Tp^{*})_{2}\right]}$ is $217.6\mathrm{\,cm^{-1}}$ away from the ground doublet and hence the investigated compound is perfectly in the scheme of an effective two level system on which we model the theory. 

This typical behavior of two maxima in $\chipp$ under a temperature variation is also found in some other reports from \citet{Jeletic2011} (see Fig. 3 therein) and \citet{Peng2017a} (see Fig. 4 therein). In those reports, since the first excited doublet is not very far away from the ground doublet, the Orbach process may involve in the formation/destruction of the two maxima together with the direct, Raman, and quantum tunneling relaxation process.

\begin{figure}
\includegraphics{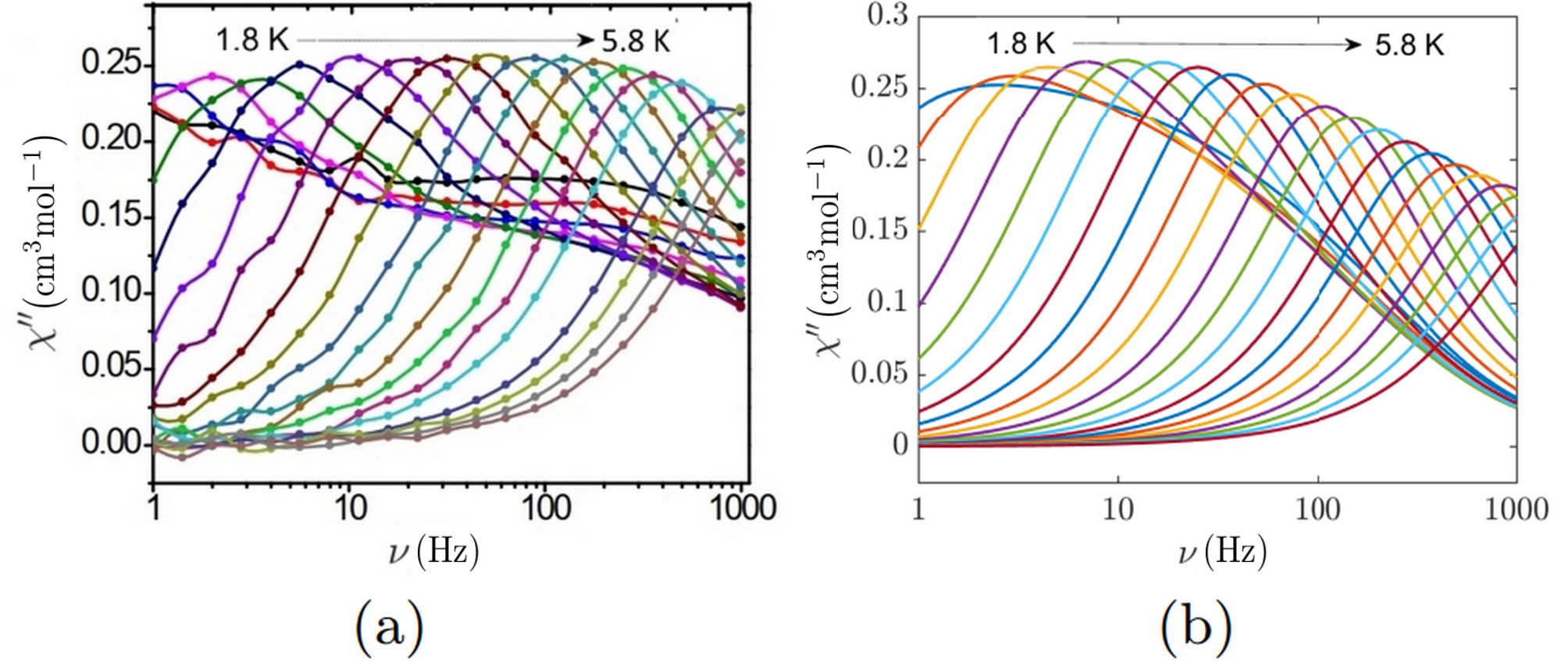}\caption{(a) Frequency dependence of the out-of-phase susceptibility of $\mathrm{\left[Co(Tp^{*})_{2}\right]}$ with 10 times magnetic site dilution under 400 Oe applied dc fields for various temperatures from \citet{Li2016}, reproduced by permission of The Royal Society of Chemistry; (b) The same quantity generated from our model using an averaging of Eq. \ref{eq:crystal out-of-phase susceptibility - no internal field} for Kramer powder sample, an addition of the second-order Raman process rate $\Gamma_{\mathrm{Raman}}=C_{\mathrm{Raman}}T^{7}$ into the expression of $\Gamma_{mm'}$, Eq. \eqref{eq:Gamma_mm'-1}, and parameters $\alpha=0$, $C_{\mathrm{Raman}}=0.025\,\mathrm{s^{-1}K^{-7}}$, $H_{n}=125\,\mathrm{Oe}$, $H_{dm}=60\,\mathrm{Oe}$, $\Gamma_{\mathrm{tn},0}=100\,\mathrm{s^{-1}}$, and $\chi_{0}=5.7\,\mathrm{cm^{3}mol^{-1}}$. \label{fig:Li2016-Dilution}}
\end{figure}

\emph{Magnetic dilution} - since the characteristic collective nuclear magnetic field $H_{n}$ results mainly from the neighbor nuclear spins of the magnetic ion(s) and of the ligands, this quantity is reduced more slowly when diluted than the characteristic dipolar field standard deviation $H_{dm}$, which mostly comes from the surrounding magnetic molecules. Accordingly, $h_{dm}\equiv H_{dm}/H_{n}$ essentially decreases upon the dilution of the SMM sample. Considering the magnitude of the dipolar field decrease is proportional to the cube of the distance between two magnetic particles, this quantity is then approximately proportional to the magnetic particle volume fraction (percentage of magnetic sites of the sample). In other words, magnetically diluting the sample by by the ratio $1:\eta$ will decrease the dipolar field distribution width around $\eta$ times. The magnetic dilution of the sample is thus able to significantly alter the appearance/disappearance of the two maxima in $\chipp$ since it substantially redistributes the contribution to $\chipp$ among slow and fast relaxation process(es). However, , whether or not it favors which specific maximum depending on many factors of which applied dc field, the characteristic collective nuclear field, and temperature are one of those. In zero applied field, the mechanism suggests only fast maximum can exists. In most cases, magnetic dilution will smear out the appearance of two maxima. This effect can be clearly seen in Fig. \ref{fig:Li2016-Dilution}a where the dynamic susceptibility of a 10 times magnetically diluted sample of $\mathrm{\left[Co(Tp^{*})_{2}\right]}$, whose undiluted sample shows two maxima in Fig. \ref{fig:Li2016}, is measured \cite{Li2016}. In order to compare the experimental results with our model, in Fig. \ref{fig:Li2016-Dilution}b, using the same parameters extracted above for the undiluted sample of $\mathrm{\left[Co(Tp^{*})_{2}\right]}$ except $H_{dm}=60\,\mathrm{Oe}$ (10 times dilution), we also plot a corresponding $\chipp$. As can be seen, the theoretical plot also shows only one maximum in $\chipp$ and possesses similar qualitative behavior as the experimental one. However, there are some noticeable difference in the absolute value of $\chipp$ at high frequency domain. Besides the possible reason of the omission of the direct process from our calculation, the difference might also come from the fact that at low concentration of the magnetic sites, the mean field approximation we have used with the dipolar field Gaussian distribution broken. 

\section{Discussions}

In this work, the appearance/disappearance of two maxima in the dynamic susceptibility of SMM samples has been investigated. Within a minimum two-level model with two relaxation processes, quantum tunneling of magnetization and direct process, we have demonstrate that this appearance/disappearance  1) cannot result solely from an averaging over the orientation of applied dc field w.r.t. the molecular frame in a polycrystalline sample (``powder EPR'' effect); 2) is due to the existence of a sufficiently wide intermolecular dipolar field distribution in the sample. Particularly, in combination with the applied dc field, the dipolar field effectively creates two distinct domains of the total magnetic field  where one relaxation process dominates the other. Two maxima in the out-of-phase susceptibility $\chipp$ will occur as two conditions are satisfied: 1) the frequency locations of the prospective maxima of the out-of-phase susceptibility $\chipp$ in the dipolar field distribution over the first and second domain are far apart; 2) the corresponding value of $\chipp$ at these frequencies are of the same order of magnitude. 

It is important to mention that although so far we have worked with a specific form of the quantum tunneling rate, Eq. \eqref{eq:Gamma_mm'-2}, the mechanism of the phenomenon is hardly dependent on this choice due to its general explained nature, provided that those two above conditions, which are quite general and independent from the mechanism of quantum tunneling, are met. 

In general, other relaxation processes may also be involved in the expression of $\chipp$. Therefore, it is interesting to know how these relaxation processes affect the appearance/disappearance of two maxima in $\chipp$. Technically, the involvement of the Raman process can be trivially done by adding the Raman relaxation rate to the expression of the total relaxation rate, Eq. \eqref{eq:Gamma_mm'-1}. The involvement of the Orbach process is a little more complicated since more energy levels will take part in the relaxation process. Theoretically, the Orbach process will start contributing when the temperature is high enough for the excited states to be populated.  In the case when the Orbach process becomes dominant, given that the dipolar field distribution and the applied dc field hardly influence this relaxation process, it is quite obvious that only one maximum in $\chipp$ appears. This is likely the common scenario in the high temperature regime. At intermediate temperature, the transition from two maxima, if any, to one maximum may take place due to the combined effect of the Orbach process and the increase in the contribution of the direct process/Raman process. The same behavior as in the case of dominant Orbach process may also be expected for a dominant Raman process, provided that it is insensitive to the change of the total magnetic field at the central spin's site \cite{Abragam1970}.

As mentioned in the introduction, in the case of mononuclear SMMs and with more than two energy level populated, the intramolecular mechanism \cite{Ho2016} may also play a role in the appearance of two maxima. Whether or not this mechanism or the intermolecular one proposed here mediates the phenomenon depends on details of the multiplet spectrum. For polynuclear SMMs, the same scenario may occur but instead of two, now three mechanisms may be involved. A situation when three maxima (or more) appear in the out-of-phase susceptibility of a polynuclear sample might occur as well. For example, one maximum may originate from the relaxation pathway of the first kind of magnetic ions in the SMM compound while two others result from another kind of the magnetic ions. In fact, this three maxima in $\chipp$ were also recently observed in some polynuclear systems \cite{Dolai2015,Yoshida2017,Funes2016}.

One should also comment on the appearance/disappearance of the second maximum in the out-of-phase susceptibility in some systems under a very strong applied magnetic field \cite{Rinehart2010,Habib2015,Boca2017}. In these systems,  the first excited (quasi-) doublet, either \emph{ab initio} computed \cite{Rinehart2010,Baldovi2013a} or from fitting the relaxation rate with Orbach process \cite{Habib2015,Boca2017}, is reckoned to be quite large comparing to the investigated temperature. Hence, these can be considered as two-level system like the one modeled here.  Since at a very strong applied field, the assumption of a white Gaussian distribution of the \emph{static} dipolar fields is less justified, the mechanism looks less realistic. A more general theory taking into account the dynamic effects of the intermolecular dipolar interaction is thus needed and will be studied in the future. 

In summary, we have proposed an intermolecular mechanism for the appearance/disappearance of two maxima in the out-of-phase susceptibility observed in SMMs under zero or weak applied dc field, which can be seen as a complement to the distinct relaxation pathways interpretation \cite{Blagg2013b,Zadrozny2011a,Hewitt2010a,Lin2009a,Guo2011d,Amjad2016,Guo2011a,Guo2010,Hewitt2009}(in polynuclear SMMs) and the intramolecular mechanism for the same phenomenon\cite{Ho2016}. The distinguishing feature of this mechanism is that it develops within only one single ground doublet. Via a simple microscopic model of two-level system with a minimum two relaxation processes involved, we proved that this phenomenon arises due to the existence of a sufficiently wide dipolar field distribution in the SMM samples. The mechanism is operative for both single-crystal and polycrystalline samples and applicable for multiple-level systems involving many relaxation processes. 
\begin{acknowledgments}
L. T. A. H. would like to acknowledge financial support from the Fonds Wetenschappelijk Onderzoek - Vlaanderen (FWO). 
\end{acknowledgments}

\end{document}